\documentclass[acmtog,nonacm]{acmart}
\acmSubmissionID{357}
\usepackage{multirow}
\usepackage{booktabs} %
\usepackage{natbib}
\usepackage{graphicx}
\usepackage{float}
\usepackage{enumitem}
\usepackage{bbding}
\usepackage{soul}
\usepackage{bm}
\usepackage{xspace}
\usepackage{array}
\usepackage{multirow}
\usepackage{overpic}

\usepackage{ulem}
\usepackage{lipsum}
\usepackage{natbib}

\usepackage[ruled]{algorithm2e} %

\SetAlFnt{\small}
\SetAlCapFnt{\small}
\SetAlCapNameFnt{\small}
\SetAlCapHSkip{0pt}

\begin{document}
\title{BANG: Dividing 3D Assets via Generative Exploded Dynamics}

\author{Longwen Zhang}
\orcid{0000-0001-8508-3359}
\affiliation{%
 \institution{ShanghaiTech University}
 \city{Shanghai}
 \country{China}}
\affiliation{%
 \institution{Deemos Technology Co., Ltd.}
 \city{Shanghai}
 \country{China}
}
\email{zhanglw2@shanghaitech.edu.cn}

\author{Qixuan Zhang}
\orcid{0000-0002-4837-7152}
\affiliation{%
 \institution{ShanghaiTech University}
 \city{Shanghai}
 \country{China}}
\affiliation{%
 \institution{Deemos Technology Co., Ltd.}
 \city{Shanghai}
 \country{China}
}
\email{zhangqx1@shanghaitech.edu.cn}

\author{Haoran Jiang}
\orcid{0009-0006-9673-8545}
\affiliation{%
 \institution{ShanghaiTech University}
 \city{Shanghai}
 \country{China}}
\affiliation{%
 \institution{Deemos Technology Co., Ltd.}
 \city{Shanghai}
 \country{China}
}
\email{jianghr2024@shanghaitech.edu.cn}

\author{Yinuo Bai}
\orcid{0009-0000-3434-212X}
\affiliation{%
 \institution{ShanghaiTech University}
 \city{Shanghai}
 \country{China}}
\affiliation{%
 \institution{Deemos Technology Co., Ltd.}
 \city{Shanghai}
 \country{China}
}
\email{baiyn2022@shanghaitech.edu.cn}

\author{Wei Yang}
\orcid{0000-0002-1189-1254}
\affiliation{%
 \institution{Huazhong University of Science and Technology}
 \city{Wuhan}
 \country{China}}
\email{weiyangcs@hust.edu.cn}

\author{Lan Xu}\authornote{Corresponding author.}
\orcid{0000-0002-8807-7787}
\affiliation{%
 \institution{ShanghaiTech University}
 \city{Shanghai}
 \country{China}}
\email{xulan1@shanghaitech.edu.cn}

\author{Jingyi Yu}\authornotemark[1]
\orcid{0000-0002-8580-0036}
\affiliation{%
 \institution{ShanghaiTech University}
 \city{Shanghai}
 \country{China}}
\email{yujingyi@shanghaitech.edu.cn}

\newcommand{\eqtr}{\mathrel{\raisebox{-0.1ex}{%
\scalebox{0.8}[0.6]{$\vartriangle$}}}}

\begin{abstract}
3D creation has always been a unique human strength, driven by our ability to deconstruct and reassemble objects using our eyes, mind and hand. However, current 3D design tools struggle to replicate this natural process, requiring considerable artistic expertise and manual labor. This paper introduces BANG, a novel generative approach that bridges 3D generation and reasoning, allowing for intuitive and flexible part-level decomposition of 3D objects. At the heart of BANG is "Generative Exploded Dynamics", which creates a smooth sequence of exploded states for an input geometry, progressively separating parts while preserving their geometric and semantic coherence.
BANG utilizes a pre-trained large-scale latent diffusion model, fine-tuned for exploded dynamics with a lightweight exploded view adapter, allowing precise control over the decomposition process. It also incorporates a temporal attention module to ensure smooth transitions and consistency across time. BANG enhances control with spatial prompts, such as bounding boxes and surface regions, enabling users to specify which parts to decompose and how. This interaction can be extended with multimodal models like GPT-4, enabling 2D-to-3D manipulations for more intuitive and creative workflows.
The capabilities of BANG extend to generating detailed part-level geometry, associating parts with functional descriptions, and facilitating component-aware 3D creation and manufacturing workflows. Additionally, BANG offers applications in 3D printing, where separable parts are generated for easy printing and reassembly. In essence, BANG enables seamless transformation from imaginative concepts to detailed 3D assets, offering a new perspective on creation that resonates with human intuition.
\end{abstract}

\ccsdesc[500]{Computing methodologies~Artificial intelligence}

\keywords{Generative Exploded Dynamics, Part-Level 3D Generation, 3D Asset Generation}

\begin{teaserfigure}
    \setlength{\abovecaptionskip}{3pt}
    \centering
    \includegraphics[width=1\textwidth,trim=0 0 0 0,clip]{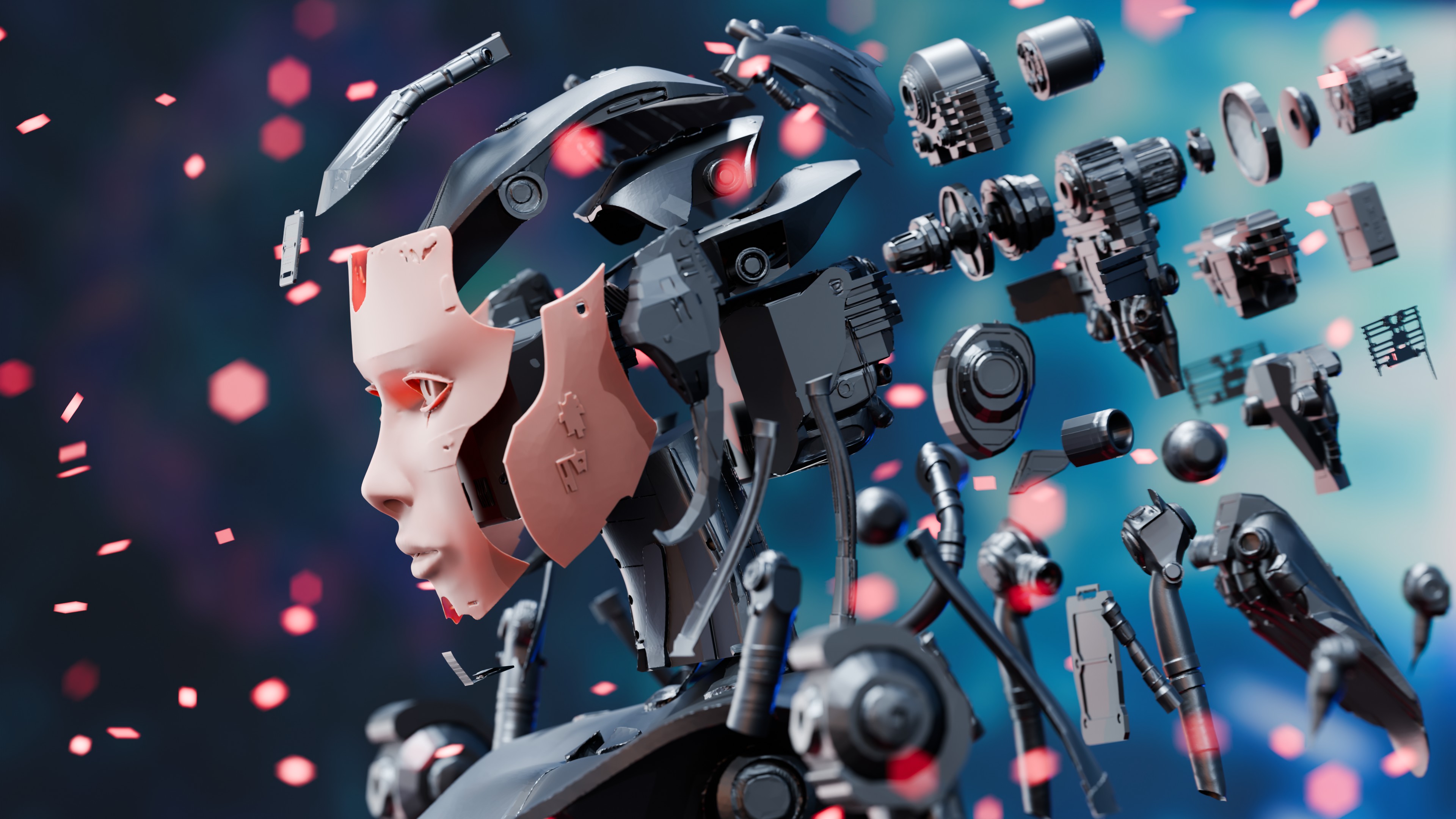}
    \caption{
{
An exploded view, generated and enhanced by our framework \textit{BANG}, of a futuristic mechanical humanoid where the fusion of organic form and mechanical precision is laid bare.
Each component of the humanoid is generated by recursively exploding its parent component using Generative Exploded Dynamics (Sec.~\ref{sec:GED}) and enhanced through Per-part Geometric Details Enhancement (Sec.~\ref{sec:partregen}). This process is conducted iteratively to create the final exploded view, which is rendered using Blender~\cite{blender}.
}
    }
    \label{fig:teaser}
\end{teaserfigure}

\maketitle

\section{Introduction}

Three-dimensional (3D) creation begins with our innate ability to understand the world around us in terms of parts and how they fit together. As children, we learned this through play—stacking stones to build majestic castles or dismantling toys like model cars and wind-up robots to explore their inner structures. Through deconstruction and recreation, we grasped the complexity of objects and experienced the joy of creation. This component-based 3D creation extends far beyond childhood and has profoundly influenced fields like computer graphics, industrial design, films, and games. However, current 3D creation tools often fail to mimic this natural ability to break down and reassemble objects. The process of decomposing and adjusting parts requires substantial artistic expertise and tedious manual effort. An ideal tool should integrate both understanding and generation at the component level, effortlessly transforming our innate creativity into tangible, interactive 3D objects.

Recent progress in Generative AI and large models has highlighted the immense potential of bridging the generation and understanding capabilities.  
In the 2D image modality, tools like DALL-E 3~\cite{openai2023dalle3}, leveraging advancements in large language models like the GPT-4 family~\cite{achiam2023gpt4}, showcase the potential of combining generative and reasoning capabilities to transform text into compelling visuals.
In the text modality, the huge breakthrough of the ``next-token prediction'' in large-language models (LLMs) has exemplified the principle that a successful way to understand is through generation.
However, unlike the image/text modalities, the 3D domain, especially for object-level content, has taken a distinct developmental path. This divergence lies in the subtle disconnection between 3D generation and reasoning. Over the past two years, 3D generation has made remarkable progress, evolving from early distillation techniques~\cite{poole2023dreamfusion}, to multi-view methods~\cite{long2024wonder3d,shi2024mvdream} and more recently to 3D native ones~\cite{zhang20233dshape2vecset,xiang2024trellis}. Yet, current mainstream approaches predominantly focus on generating entire objects in one piece, lacking the component-based capability for flexible manipulation and detailed design. On the other hand, 3D understanding has advanced in component-level analysis. For instance, some methods~\cite{yang2024sampart3d,zhou2024pointsam} provide instance-level part segmentation, while others~\cite{qi2025shapellm,xu2025pointllm} integrate 3D features with LLMs to enable dialogue-driven reasoning. However, they often focus on the visible outer surface, neglecting the occluded internal structure, and hence struggle to establish spatial and semantic interconnections within the 3D object.
In a nutshell, a more natural approach is needed to bridge 3D generation and reasoning—one that mirrors how we intuitively understand and create by dividing and assembling objects as children, aligning with the idea that understanding is achieved through generation.

Inspired by the Big Bang Theory, where a singularity bursts into stars, planets, and life, we introduce \textit{BANG}—a generative approach that dynamically divides complex 3D assets into interpretable parts through a smooth, consistent ``exploding'' process. Much like how the universe transitioned from a unified state to a dispersed one, BANG allows 3D objects to be divided and reassembled in a way that preserves both structure and coherence. Just as children naturally learn by taking apart and reassembling their toys, BANG deconstructs in generation and reconstructs in understanding. BANG allows for high-quality 3D decomposition, generation, and enhancement while seamlessly integrating part-level analysis, bridging our 3D concept imagination into digital creation.

The core of BANG lies in a novel design called ``Generative Exploded Dynamics'', which transforms an input geometry into a continuous sequence of exploded states through a smooth radial explosion process. Each intermediate state is represented as a single mesh, where constituent parts progressively separate while preserving semantic and geometric consistency. It culminates in a fully divided state, akin to the exploded view commonly used for asset visualization.
Unlike static surface segmentation, generating exploded dynamics progressively separates parts over time, enabling the model to uncover latent volumetric structures and internal boundaries. This dynamic separation process naturally captures geometric and semantic dependencies that are otherwise difficult to infer.
To achieve this, we adopt a diffusion-based generative model with the ``pretrain-then-adaptation'' paradigm. We first pre-train a large-scale latent diffusion model on static 3D geometry with  neural field representation based on 3DShape2VecSet~\cite{zhang20233dshape2vecset}, leveraging high-quality geometry priors. Then, we fine-tune the base model for exploded dynamics, using a carefully designed dataset with rich part-level assembly structures. Specifically, we propose a light-weight exploded view adapter to condition the base model on input geometry and timestamps, enabling precise and smooth decomposition. We also adopt a temporal attention module to enhance smooth transitions and maintain semantic and geometric consistency across timestamps.
Beyond generating divided parts, we further utilize part-aware trajectory tracking compatible with the neural field representation. It associates the components back to the original mesh for accurate reassembly and preserves part semantics and spatial coherence.

Achieving control over object decomposition is crucial for innovative and efficient 3D creative workflows. To enhance controllability, we further explore two kinds of cross-attention-based spatial prompts for BANG: bounding boxes and surface regions. Bounding boxes can specify volumetric regions even for geometries without internal structures, while surface regions enable precisely isolating and manipulating detailed areas on the object's surface. Additionally, BANG inherently preserves geometric and spatial semantics. Thus, we decode and align its 3D features with 2D feature extractors and collaborate with multimodal models (e.g., DINOv2~\cite{oquab2023dinov2} and Florence-2~\cite{xiao2024florence2}). This enables intuitive 2D-to-3D interactions where one can specify object regions directly on 2D rendered views or sketches for controllable generation.

The strength of BANG lies in its ability to transform complex 3D assets into detailed, interpretable parts. BANG allows users to generate, decompose, and reassemble objects from simple text or image inputs,  enhancing geometric details at the part level. Integrated with large multi-modal models, BANG enables interactive dialogues for part-level 3D analysis and creation, while also supporting 3D printing and assembly with an engaging, hands-on creation experience. Through BANG, the process of creating and understanding 3D objects becomes as intuitive and joyful as assembling a puzzle, piece by piece. As Feynman once said, “What I cannot create, I do not understand.” BANG brings this idea to life, turning imagination into reality.

\section{Related Work}

\subsection{3D Structural Understanding}

Understanding the intricate structure of 3D objects and providing functional and semantic analysis of their constituent parts facilitates advanced operations of 3D assets. Here we primarily review the methods for part-level semantic segmentation and those integrating large models for dialogue-driven reasoning.

\paragraph{Part Segmentation.}

Current approaches for 3D part segmentation largely focus on exploring network architectures for point cloud or mesh of outer surface~\cite{guo2015dcnn,xu2017cnn,qi2017pointnet,qi2017pointnet++,li2018pointcnn,zhao2021pointtransformer,qian2022pointnext,ma2022pointmlp,wu2022ptv2,wu2024ptv3}. They heavily rely on labeled datasets such as PartNet~\cite{mo2019partnet}, which, while valuable, are limited in size and scope, often encompassing specific categories like furniture. 
To enhance generalization ability, recent zero-shot and open-vocabulary approaches~\cite{abdelreheem2023satr,liu2023partslip,tang2024segmentanymesh,thai2025_3x2,umam2024partdistill,zhong2024meshsegmenter,zhou2023partslip++,yang2024sampart3d,cen2023sa3d,yang2023sam3d,takmaz2023openmask3d,jatavallabhula2023conceptfusion,liu2024part123,zhang2022pointclip,zhu2023pointclipv2} leverage pretrained large-scale vision models, i.e., CLIP~\cite{radford2021clip}, DINO~\cite{caron2021dino,oquab2023dinov2}, GLIP~\cite{li2022GLIP,zhang2022glipv2}, and SAM~\cite{kirillov2023sam,ravi2024sam2}. They render 3D objects into 2D images to apply these vision models, hence inherently limiting segmentation to visible surfaces and ignoring the internal components.

\paragraph{Multi-modality Analysis.}
Recent methods focus on multi-modality analysis of 3D objects. They have led to the development of scalable 3D encoders that align 3D features with those from text and image encoders. These encoders facilitate a range of tasks, i.e., 3D feature extraction~\cite{xue2023ulip,xue2024ulip2,liu2024openshape,zhang2023i2pmae, zhou2023uni3d} and descriptive question and answer (Q\&A) systems combined with LLMs~\cite{yin2023shapegpt,qi2025shapellm,qi2024gpt4point,fei2024kestrel,qi2025gpt4scene,tang2024minigpt3d,hong20233dllm,xu2025pointllm,ma2024llms}. These models, trained on extensive datasets such as Objaverse~\cite{deitke2023objaverse}, offer a comprehensive understanding of 3D objects by capturing both their geometric features and semantic attributes. However, they focus on surface geometry, overlooking the essential aspect of internal volumetric structural understanding. 

Differently, our BANG approach effectively displaces parts and models interior components through generative exploded dynamics, surpassing surface-level methods for purely 3D  understanding. Our isolated parts improve generative mesh quality and semantic consistency for precise manipulation. It can serve as a plausible precursor for 3D segmenting anything from outer to inner and is compatible with LLMs such as GPT-4 family to facilitate component-level descriptive and query capabilities.

\begin{figure*}
    \centering
    \includegraphics[width=1\linewidth]{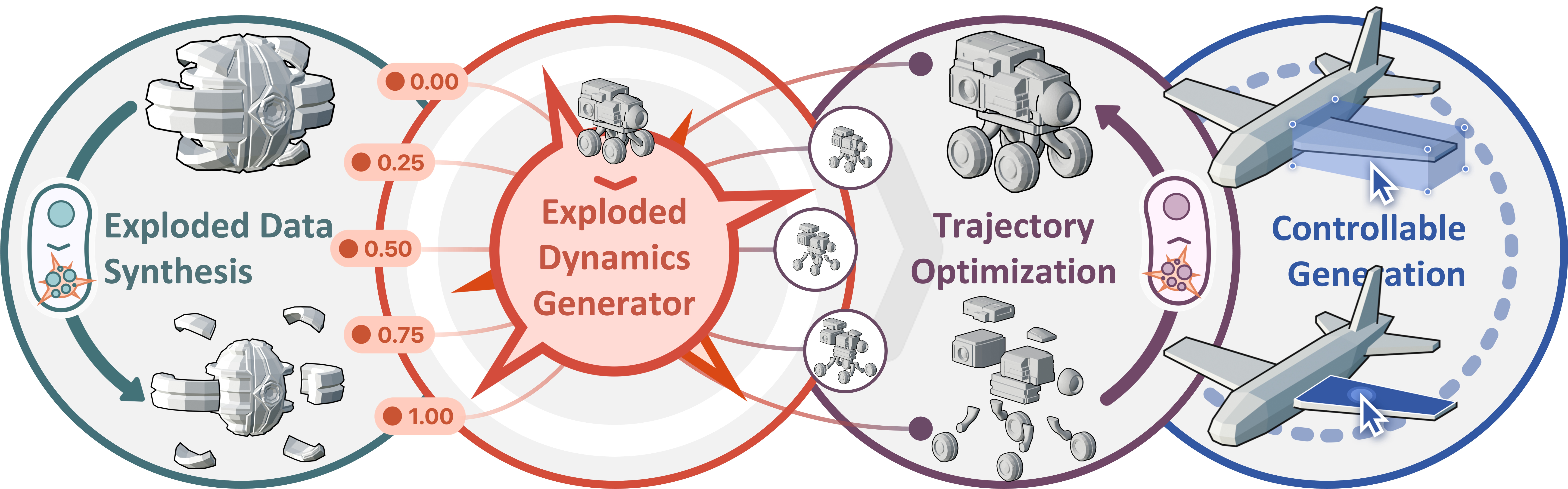}
    \caption{
The overview illustrates the proposed framework for \textit{Generative Exploded Dynamics}. The pipeline consists of four stages: 
Exploded Data Synthesis generates the training data (Section.~\ref{sec:dataset}). 
Exploded Dynamics Generator produces the exploded dynamics based on the input geometry (Section.~\ref{sec:architecture}). 
Trajectory Optimization refines the trajectories of the exploded parts, ensuring smooth reassembly of the components (Section.~\ref{sec:tracking}). 
Finally, Controllable Generation allows users to interactively control and refine the explosion by conditions (Section.~\ref{sec:control}).
    }
    \label{fig:overview}
\end{figure*}

\subsection{3D Generation}
Here, we systematically review recent progress in 3D object generation, including those generating entire objects as a whole through 2D lifting or using 3D native representation, as well as those focusing on part-aware generation.

\paragraph{2D Lifting.}
Pioneering works like DreamFusion~\cite{poole2023dreamfusion} introduce Score Distillation Sampling (SDS) and optimize underlying geometric representations using 2D diffusion priors, while Zero-1-to-3~\cite{liu2023zero1to3} generates multi-view images from a single image input. Building upon them, a significant volume of subsequent research has explored this 2D-to-3D lifting paradigm~\cite{lin2023magic3d,qian2023magic123,chen2024GSGEN,yi2024gaussiandreamer,huang2023dreamtime,wang2023sjc,xu2023neurallift,raj2023dreambooth3d,gu2023nerfdiff,xiang2023ivid,wang2024prolificdreamer,melas2023realfusion,tang2023makeit3d,watson20223dim,liu2024one2345}. A key direction is to improve multi-view consistency, achieving more coherent and accurate 3D reconstruction~\cite{chan2023genvs,tang2025mvdiffusion++,liu2024one2345++,qiu2024richdreamer,long2024wonder3d,liu2024syncdreamer,shi2024mvdream,shi2023zero123++,li2023instant3d,gao2024cat3d,chen2024v3d}.
Besides, researchers have also explored the creation of composite objects and entire scenes \cite{cohen2023setthescene,li2024focaldreamer,po2024comp3d,yan2024dreamdissector,epstein2024d3ll,wang2023luciddreaming,vilesov2023cg3d,han2024reparo,chen2025comboverse,yan2024phycage}. These methods usually leverage the understanding of object arrangements embedded in the image-based generative models and employ differentiable rendering to optimize the placements of individual objects within a scene.

\paragraph{3D Native Generation.}
Another direction involves training the generative models directly using extensive 3D data of diverse shapes and styles. Exemplified by 3DShape2VecSet~\cite{zhang20233dshape2vecset}, CLAY~\cite{zhang2024clay}, and TRELLIS~\cite{xiang2024trellis}, these 3D native methods produces impressive geometry and appearance~\cite{deng2024detailgen3d,jun2023shape,nichol2022pointe,ren2024xcube,wu2024direct3d,li2024craftsman,Zheng2023LAS}. Further explorations, exemplified by PolyGen~\cite{nash2020polygen}, MeshGPT~\cite{siddiqui2024meshgpt}, and Meshtron~\cite{hao2024meshtron}, adopt an autoregressive generation approach for mesh faces~\cite{chen2024meshxl,weng2024pivotmesh,chen2024meshanything,chen2024meshanything2,tang2024edgerunner,weng2024scaling}.
Besides, a related area of research focuses on 3D CAD generation, which rely on structured CAD representations with explicit awareness of components and primitives~\cite{li2022free2cad,you2024img2cad,alam2024gencad,khan2024textcad,badagabettu2024query2cad,uy2022point2cyl,dupont2025transcad,xu2024cadmllm}.

\paragraph{Part-aware Generation.}
While the above generative models excel at producing unified meshes, their lack of explicit part separation limits component-level editing and interaction. Part-level generation addresses this limitation which requires not only the creation of individual parts but also their coherent assembly into complete objects.  Early studies~\cite{gao2019sdmnet,wu2019sagnet,mo2019structurenet,wu2020pqnet,petrov2023anise,hertz2022spaghetti} focused on encoding and decoding part geometries and positions, laying the groundwork for part-aware generation. Subsequently, approaches~\cite{nakayama2023difffacto,koo2023salad} have demonstrated the potential for fine-grained part generation using diffusion models and part-specific latent representations, yet within relatively smaller and specialized datasets such as ShapeNet~\cite{chang2015shapenet} and PartNet~\cite{mo2019partnet}. The recent PartGen~\cite{chen2024partgen} handles occlusion through a two-stage process: first, producing artist-inspired part segmentation through multi-view synthesis, followed by generating the detailed 3D shapes for each part. 

In stark contrast, our BANG approach natively decomposes objects into meaningful parts and ensures their coherent reassembly through innovative exploded dynamics. Unlike prior methods that rely on multi-view segmentations or two-stage processes, it inherently encodes structural understanding within a unified large-scale generative paradigm, offering flexibility and precision for both creation and downstream applications.

\subsection{4D Generation and Exploded View}

Our BANG produces a special dynamic sequence of exploded 3D geometries where constituent parts of the original mesh progressively separate. Hence, it partially shares common insights with those approaches about generating dynamic 4D objects and traditional exploded views.
Specifically, recent efforts generate dynamic objects or scenes using NeRF or Gaussian representations~\cite{yin20234dgen,rahamim2024bo2l, liang2024diffusion4d, singer2023mav3d, bahmani20244dfy, zhao2023animate124, jiang2023consistent4d, ren2023dreamgaussian4d, zeng2025stag4d, pan2024efficient4d}. Some of them have adapted 3D generative models to handle 4D temporal sequences \cite{cao2024motion2vecsets,zhang2024dnf,erkocc2023hyperdiffusion} using similar strategies in BANG, i.e., temporal attention.
On the other hand, traditional exploded views separate the components of a 3D object to expose its internal structure, providing an intuitive way to perceive complex 3D architectures. Existing work on exploded view generation has predominantly concentrated on 2D representations~\cite{li2008explodedgeneration,karpenko2010exploded,li2004interactive,bruckner2006explodedvolume,Shao2021explodedprojection}. Exploded views in 3D have been largely overlooked despite their intuitive appeal.
Differently, our approach introduces a native method for generative exploded dynamics that integrates 3D part-level decomposition within a large-scale generative framework. This not only offers a novel approach for part-aware 3D generation but also open up new possibilities for 3D creation workflows.

\section{Generative Exploded Dynamics}
\label{sec:GED}

Differently, our BANG approach dynamically divides complex 3D assets into interpretable part-level structures, to deconstruct in generation and reconstruct in understanding. 
As illustrated in Fig.~\ref{fig:overview}, the core of BANG is a novel design called \textit{Generative Exploded Dynamics}. Within a conditioning-generative paradigm, it simulates a smooth and radial ``explosion'' process, transitioning a complete and assembled geometry into its constituent parts. Crucially, each intermediate exploded state preserves part-level geometric and semantic consistency, ensuring a sequence of meaningful decomposition. As a result, our framework encapsulates sophisticated structural insights to facilitate both the fidelity and controllability of 3D geometry generation and analysis.

For clarity of exposition, we first detail the architecture design and training strategy, explaining how geometry is encoded, interpreted, and ultimately decomposed into continuous exploded dynamics (Sec.~\ref{sec:architecture}). We then describe our data preprocessing to facilitate robust training (Sec.~\ref{sec:dataset}). 
Finally, we introduce our post-generation trajectory tracking procedure (Sec.~\ref{sec:tracking}), which is applied to geometry sequences generated by our model, ensuring stable part-wise transitions, semantic consistency, and accurate reassembly.

\begin{figure}[t]
    \centering
    \includegraphics[width=1\linewidth]{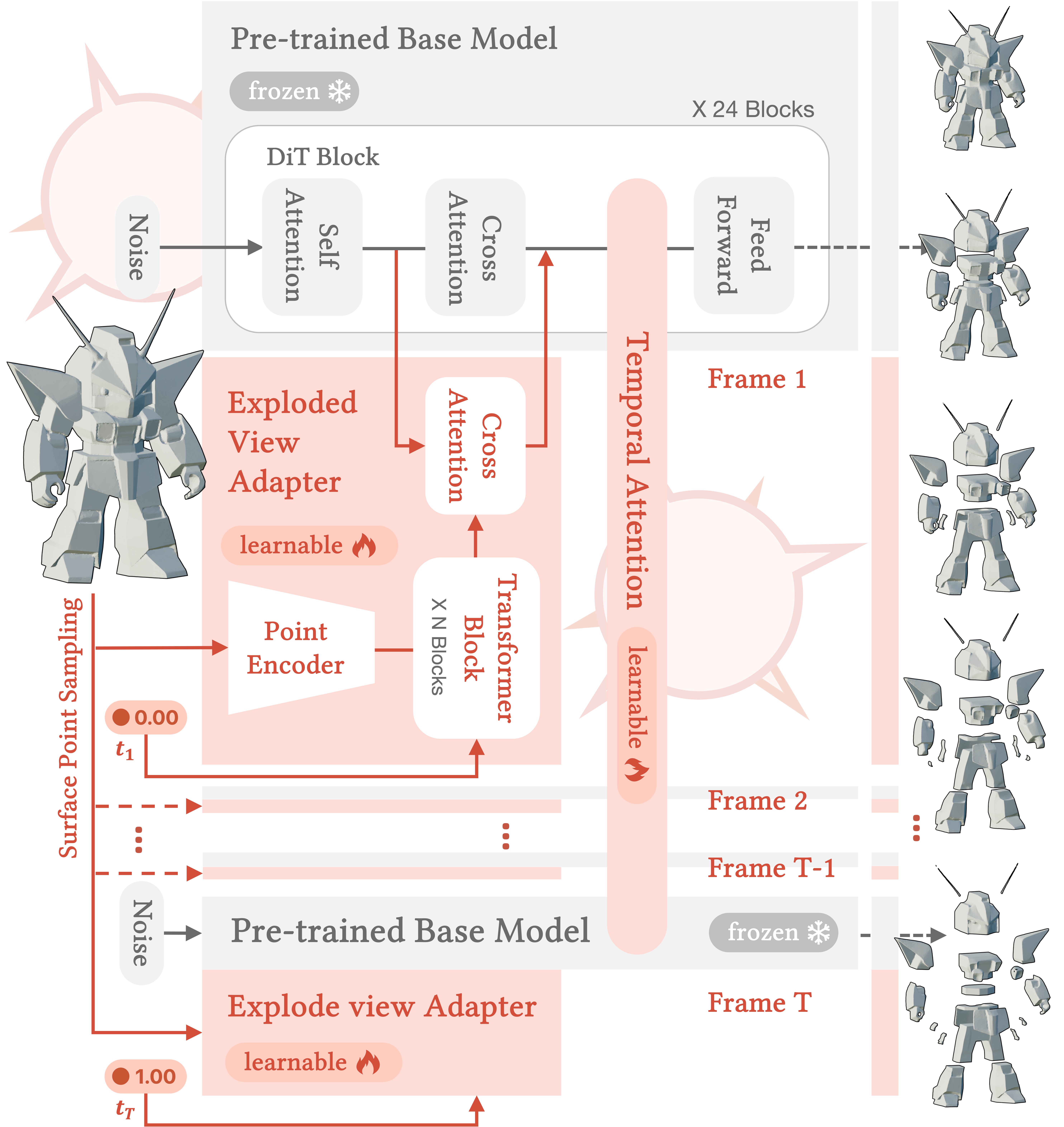}

    \caption{
The architecture of the base generative model and adaptation modules in our \textit{Generative Exploded Dynamics} framework. The gray blocks represent the pretrained base model, which is a transformer-based latent diffusion model, and remains frozen after pretraining. The red blocks include the exploded view adapter and temporal attention module, which are learnable during the exploded dynamics training phase. During inference, input geometry, along with a target time sequence $\{t\}$, is fed into the exploded view adapter. Temporal attention ensures that the entire diffusion model outputs a continuous, smoothed geometry sequence in one pass.
    }
    \label{fig:architecture}
\end{figure}

\subsection{Exploded Dynamics Generation Model}
\label{sec:architecture}
We adopt a diffusion-based generative model to produce a series of meshes from an assembled state to a smoothly exploded configuration.
Formally, given an input geometry $\mathcal{M}$ and a time series $t \in \{t_1,\dots,t_T\}$ as conditions, it generates the corresponding watertight mesh sequence $\{\mathcal{M}_t\}$. In $\{\mathcal{M}_t\}$, all the constituent parts in the original mesh are naturally and continuously transited from a fully assembled state ($t=0$) to a completely divided state ($t=1$).
As shown in Fig.~\ref{fig:architecture}, we adopt a ``pretrain-then-adaptation'' scheme. We first pretrain a large-scale 3D generative model for high-quality and static geometry modeling similar to previous methods~\cite{zhang20233dshape2vecset,zhang2024clay}. Next, we fine-tune the large model into our exploded setting using a part-specific exploded-view dataset. Specifically, to achieve precise and smooth part-level decomposition, we propose an \textit{Exploded View Adapter} that conditions the model on input geometry and various timestamps. Additionally, we adopt a \textit{Temporal Attention Module} to ensure smooth and coherent part transition across the exploded process.
These designs collectively enhance the ability to generate part-aware dynamics with high fidelity.

\paragraph{3D Generative Model pretraining}
\label{sec:pretraining}
Similar to prior works leveraging 3DShape2VecSet representation~\cite{zhang20233dshape2vecset,zhang2024clay}, our base model consists of a geometry variational autoencoder (VAE) and a latent diffusion model (LDM). 
To encode a 3D geometry, we first sample a point cloud $\bm{X}$ from the surface of the input mesh $\mathcal{M}$. $\bm{X}$ is then transformed into a latent representation $\bm{Z} \in \mathbb{R}^{L\times C}$ by a transformer-based VAE encoder:
\begin{equation}
     \bm{Z} =\mathcal{E}(\bm{X})=\text{CrossAttn}(\text{PosEmb}(\bm{\tilde X}),\text{PosEmb}(\bm{X})),
\end{equation}
where $\bm{\tilde X}$ denotes a down-sampled version of $\bm{X}$, $L$ is the number of points in $\bm{\tilde X}$ and $C$ is the channel dimension.
Next, we apply a diffusion transformer (DiT) model $\epsilon(\bm{Z} + \epsilon_\tau, \tau)$ to learn to denoise the noisy latent $\bm{Z} + \epsilon_\tau$.
Finally, the VAE decoder $\mathcal{D}$ processes these latent codes and a list of query points $\bm{p}$ in space, outputting SDF values:
\begin{equation}
\mathcal{D}(\bm{Z},\bm{p})=\text{CrossAttn}(\text{PosEmb}(\bm{p}),\text{SelfAttn}^{24}(\bm{Z})).
\end{equation}
We adopt the pretraining scheme~\cite{zhang20233dshape2vecset,zhang2024clay} to train both the VAE and the LDM on the Objaverse dataset \cite{deitke2023objaverse}. Additionally, we enhance the pretrained model by incorporating text, image, and point cloud conditioning schemes (implementation details are provided in Sec.~\ref{sec:details}).
Our pretrained base model establishes robust geometry prior and can generate diverse 3D geometries from diverse inputs like text prompts and images.

\paragraph{Exploded View Adapter}
\label{sec:exploded_view_adapter}
We aim to adapt the above pretrained model to generate a sequence of geometries, $\mathcal{M}_t, t \in \{t_1,\dots,t_T\} $, from an arbitrary 3D geometry $\mathcal{M}$. 
These $\mathcal{M}_t$ provide a unique perspective of the original $\mathcal{M}$ by smoothly and radically ``exploding'' its constituent parts, akin to the exploded view commonly used for asset visualization.
Specifically, we freeze the pretrained base model, then inject conditional signals derived from $\mathcal{M}$ and time $\{t\}$ into it. This design minimizes data requirements by restricting training to a lightweight adapter while retaining the strong geometric priors encoded in the base model. 

The adapter begins by encoding $\mathcal{M}$ into unordered feature representations. Following the structure of the VAE encoder $\mathcal{E}$, we uniformly sample a point cloud $\bm{S}\in \mathbb{R}^{N\times 3}$ from the surface of the input mesh $\mathcal{M}$. This sampled point cloud is then embedded and processed through a cross-attention encoding module, mirroring the encoding pipeline of $\mathcal{E}$, as follows:
\begin{equation}
\label{eq:pointencoder}
     \bm{G} =\text{CrossAttn}(\text{PosEmb}(\bm{\tilde S}),\text{PosEmb}(\bm{S})),
\end{equation}
where $\bm{\tilde S}$ denotes a down-sampled version of $\bm{S}$ via farthest-point sampling (FPS). In our implementation, $N$ is set to $20480$ with a down-sampling factor of $10$.
The resulting geometry feature $\bm{G}$ is then passed through a lightweight transformer equipped with adaptive Layer Normalization (adaLN) to incorporate the time condition $t$ and the expected parts count. This process produces the conditioning feature $\bm{G}_\text{explode}$, which is fed into the diffusion backbone.
Finally, $\bm{G}_\text{explode}$ is integrated into the main DiT backbone through parallel cross-attention layers. Specifically, the input of each cross-attention layer in the DiT backbone is cross-attended with $\bm{G}_\text{explode}$, and the resulting features are added back to the output of the corresponding DiT cross-attention layer. This mechanism ensures that the conditioning information from the exploded view adapter seamlessly guides the generative process. The adapter module is trained to align with the target exploded dynamics as follows:
\begin{equation}
\epsilon(\bm{Z}_t + \epsilon_\tau, \tau, \bm{G}_\text{explode}) \rightarrow \bm{Z}_t
\end{equation}
where $\bm{Z}_t$ denotes the exploded view latent code at time $t$, encoded from $\mathcal{M}_t$ as $\bm{Z}_t = \mathcal{E}(\mathcal{M}_t)$, and $\epsilon_\tau$ represents Gaussian noise at noise step $\tau$.
As illustrated in Fig.~\ref{fig:architecture}, this modular design ensures that the adapter can be trained independently to interpret $\mathcal{M}$ and $t$ without altering the pretrained diffusion parameters. This approach simplifies the overall pipeline while preserving the broad shape priors learned from large-scale data. Once trained, the adapter directs the model to generate exploded states from the input geometry at a target time, establishing a foundation for the subsequent multi-frame or time-sequence generation.

\begin{figure}[t]
    \centering
    \includegraphics[width=1\linewidth]{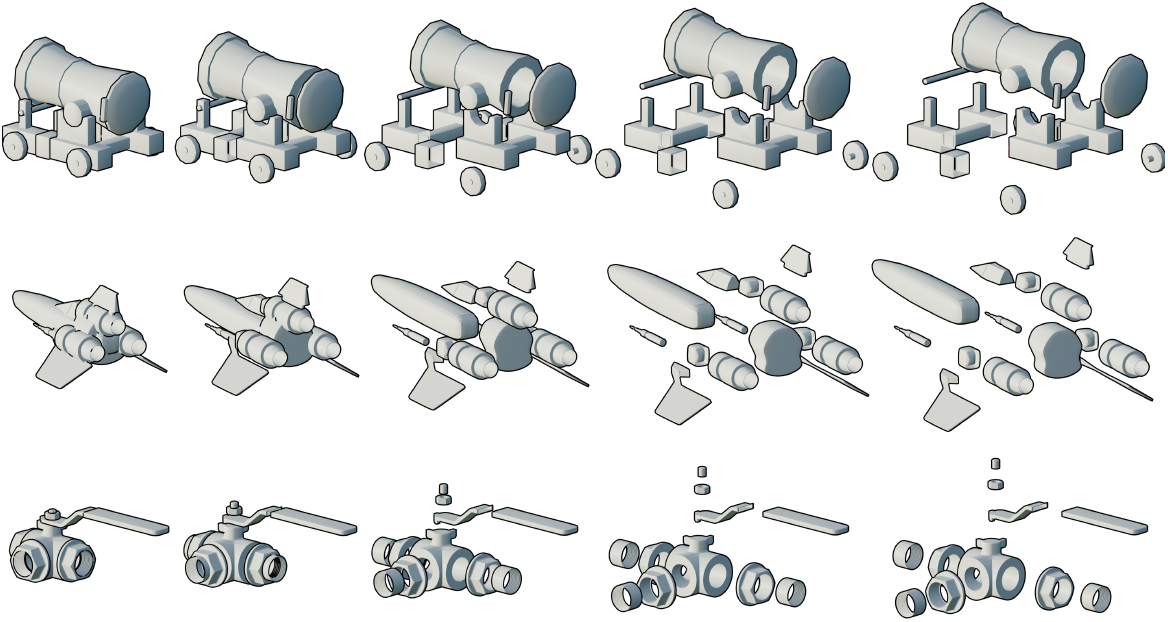}
    \caption{
Example data from our dataset illustrating synthetic exploded dynamics. The images show a cannon (top), spaceship (middle), and valve (bottom) transitioning from $t=0$ (left) to $t=1$ (right), highlighting the decomposition of each object over time.
    }
    \label{fig:datapreview}
\end{figure}

\paragraph{Temporal Attention for Smooth Exploded Sequence Generation}

To ensure smooth transitions between exploded states, we adopt a temporal attention mechanism across the exploded process. This mechanism facilitates continuity by modeling dependencies between consecutive frames, inspired by recent video diffusion models.
With the fine-tuned exploded view adapter, we extend our approach to generate a smooth exploded dynamic sequence of $T$ frames. To share contextual information across frames, we integrate a temporal attention mechanism within each DiT block. During the training of the temporal attention module, a batch of full-length exploded dynamics sequences $\{\bm{Z}_t\}, t \in \{t_1,\dots,t_T \} $ is fed into the DiT model. To enable the temporal attention module to distinguish time progression, we introduce a frame-wise time embedding, $\text{TimeEmb}(t)$, where tokens corresponding to the same frame index share the same embedding. This embedding is defined as:
\begin{equation*}
    \text{TempAttn} =\text{SelfAttn}(\bm{Z}_{t_1} \circ \text{TimeEmb}(t_1), \dots, \bm{Z}_{t_T} \circ \text{TimeEmb}(t_T)),
\end{equation*}
where $\circ$ denotes that the time embedding is only added to the query and key representations of the attention layer, similar to the Rotary Positional Embedding (RoPE) widely applied in large language models:
\begin{equation}
  \bm{q} \leftarrow \bm{q} \oplus \text{TimeEmb}(t), 
  \quad
  \bm{k} \leftarrow \bm{k} \oplus \text{TimeEmb}(t).
\end{equation}

To train the temporal attention module, we merge the token and frame dimensions into a single dimension prior to feeding the data into the temporal attention module, forming a contiguous set of $T \times L$ tokens for each instance in the batch. This transformation allows the multi-head self-attention operation to be applied across all $T \times L$ tokens, enabling the model to establish both intra-frame consistency and inter-frame transitions by allowing tokens to attend across frames.
After the temporal attention operation, the tokens are reshaped back to separate the frame dimension, and the frame dimension is then merged into the batch dimension for frame-wise generation. This design ensures that the temporal attention module captures global temporal context while maintaining the ability to generate each frame independently during subsequent stages.
This design requires only the addition of a lightweight layer for temporal coordination, which can be trained independently. It ensures seamless integration of temporal coherence into the generation process while preserving the flexibility and robustness of the underlying 3D generative model.

\begin{figure}[t]
    \centering
    \includegraphics[trim={10 0 0 0},clip,width=1\linewidth]{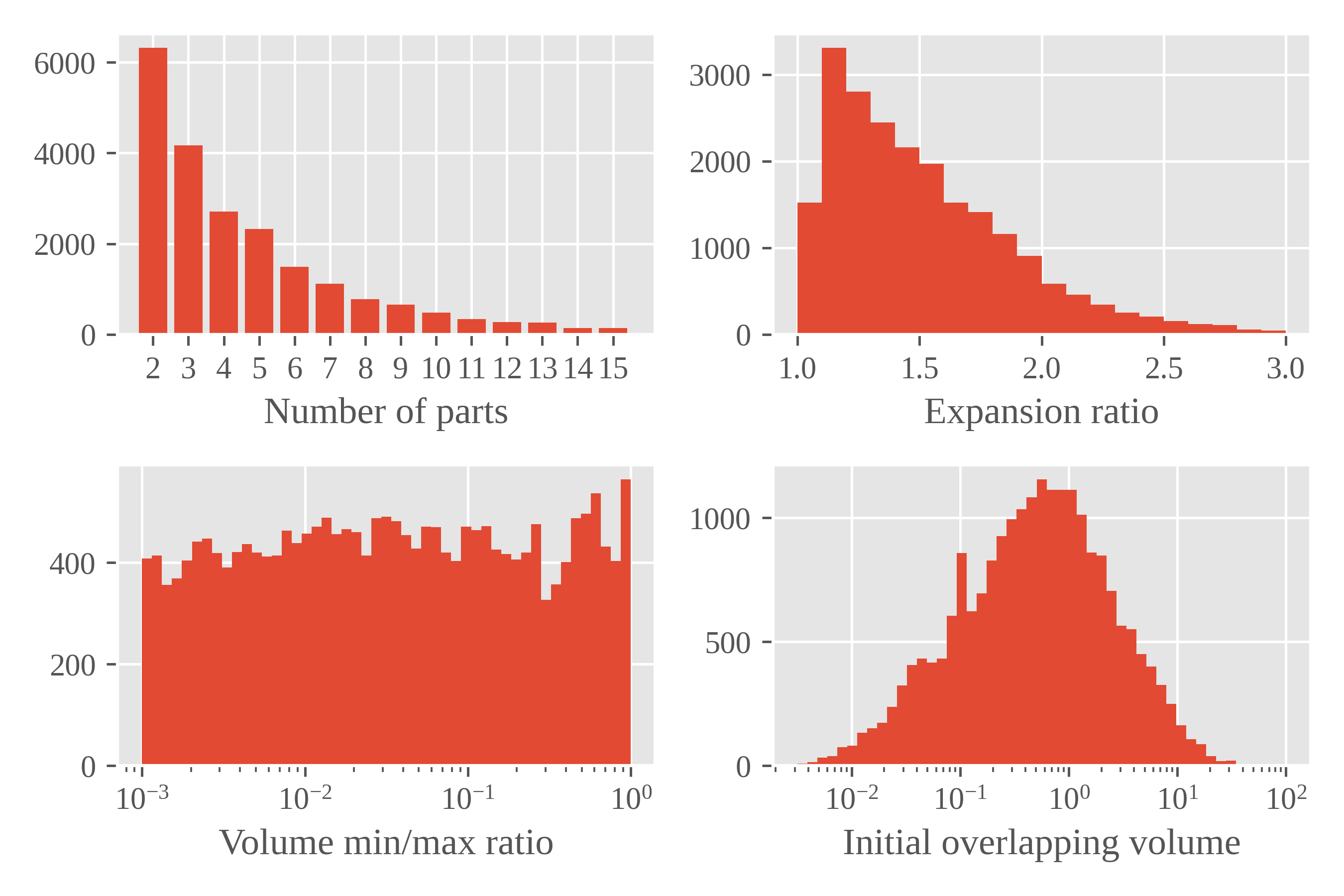}

    \caption{
Histograms illustrating the distribution of key geometric characteristics within the data used to train exploded dynamics. The plots show the distribution of: (top left) the number of parts comprising each 3D asset; (top right) the asset bounding box expansion ratio after explosion optimization, which is calculated as the maximum dimension of the bounding box after explosion divided by the maximum dimension of the bounding box at the assembled state; (bottom left) the ratio between the minimum and maximum volumes of the parts, in logarithmic scale; and (bottom right) the initial overlapping volume between parts in the assembled state, in logarithmic scale.
    }
    \label{fig:dataset}
\end{figure}

\subsection{Dataset Construction}
\label{sec:dataset}

Obtaining high-quality part-level mesh data is critical for training the generative exploded dynamics model. However, this task presents significant challenges, as most publicly available 3D assets were not designed or curated with explicit sub-component structures. Even within large repositories like Objaverse~\cite{deitke2023objaverse}, many assets are single-piece meshes, contain incomplete or poorly defined part geometries, or fail to meet quality or technical standards necessary for reliable training. To address these issues and ensure consistency in our training pipeline, we implement a rigorous filtering process for 3D assets in Objaverse, prioritizing quality over quantity to curate a robust and reliable dataset.

\paragraph{Data Filtering.} 
For assets in Objaverse, we begin by identifying assets with a component count between 2 and 30. We exclude meshes with extreme vertex counts (e.g., $<1\text{e}3$ or $>1\text{e}6$) and those containing skins intended for animation. To further ensure data quality, similar to previous work~\cite{luo2024scalable}, we conduct a thorough quality check using GPT-4~\cite{achiam2023gpt4} to identify and remove problematic meshes. 
Specifically, we render each 3D asset from multiple viewpoints and prompt GPT-4 to assess its suitability for training, filtering out scans, incomplete or unrecognizable objects, and complex scenes. 
This filtering process produces a stable subset of meshes that balance geometric detail with computational feasibility.
For the accepted meshes, GPT-4 is also used to annotate key geometric and semantic attributes, including symmetry, polygon density, and visual complexity.

\paragraph{Explosion Vector Optimization.} For each remaining mesh, we calculate the axis-aligned bounding boxes of its components and optimize a translation vector for each component to simulate a radial explosion outward. This optimization process aims to minimize collisions between bounding boxes while constraining excessive translations, ensuring the object's layout stays cohesive. It's terminated when the overlap between bounding boxes falls below a predefined small threshold. This results in a visually coherent radial explosion process. We then interpolate the translation vectors from $t=0$ (assembled) to $t=1$ (fully exploded), and sample intermediate time steps to form a smooth sequence of exploded states. To ensure consistency and simplify downstream processing, these sequential meshes are re-centered or uniformly scaled so that their overall bounding box remains within a standardized size.
If the parts remain too close or the final exploded view becomes excessively large and unrealistic, we discard the corresponding data. Finally, We record the exploded sequence, the transformations, and all relevant metadata in our dataset to ensure reproducibility and adherence to consistent standards for training and evaluation. The mesh examples of our synthetic exploded dynamics are shown in Fig.~\ref{fig:datapreview}.

\paragraph{Exploded Dynamics Dataset.}
After rigorous filtering, we curate an exploded dynamics dataset containing approximately 20k high-quality assets, with the corresponding statistics illustrated in Fig.~\ref{fig:dataset}. Although this final data set is relatively small compared to the original pool of millions of meshes, it offers precise and rich data that ensure high-quality training.
Besides, as discussed in Sec~\ref{sec:pretraining}, our method leverages a large-scale pretrained generative model, reducing the need for a dedicated dataset of exploded shapes for fine-tuning. This strategy combines the broad geometric knowledge from large-scale pretraining with the precise and unique part-level annotations in our final dataset. It enables robust part-aware generation while preserving the benefits of large-scale pretraining. In practice, this dataset is crucial for guiding the generative model in accurately decomposing parts and transitioning smoothly between assembled and exploded states. By prioritizing consistency and correctness, we reduce the noise that could hinder the convergence of our exploded dynamics generation.

\subsection{Part Trajectory Tracking}
\label{sec:tracking}

Given a generated exploded dynamics sequence, each state is represented as a single mesh composed of multiple disconnected components. To enable accurate part-level understanding, we must establish consistent correspondences between parts in the fully exploded state and their counterparts in the original geometry. This tracking process is applied after generation and allows us to follow individual components throughout the explosion sequence, preserving semantic meaning and geometric consistency across frames. This not only enhances structural understanding but also unlocks versatile editing capabilities and seamless integration with downstream applications.

\begin{figure}
    \centering
    \includegraphics[width=1\linewidth]{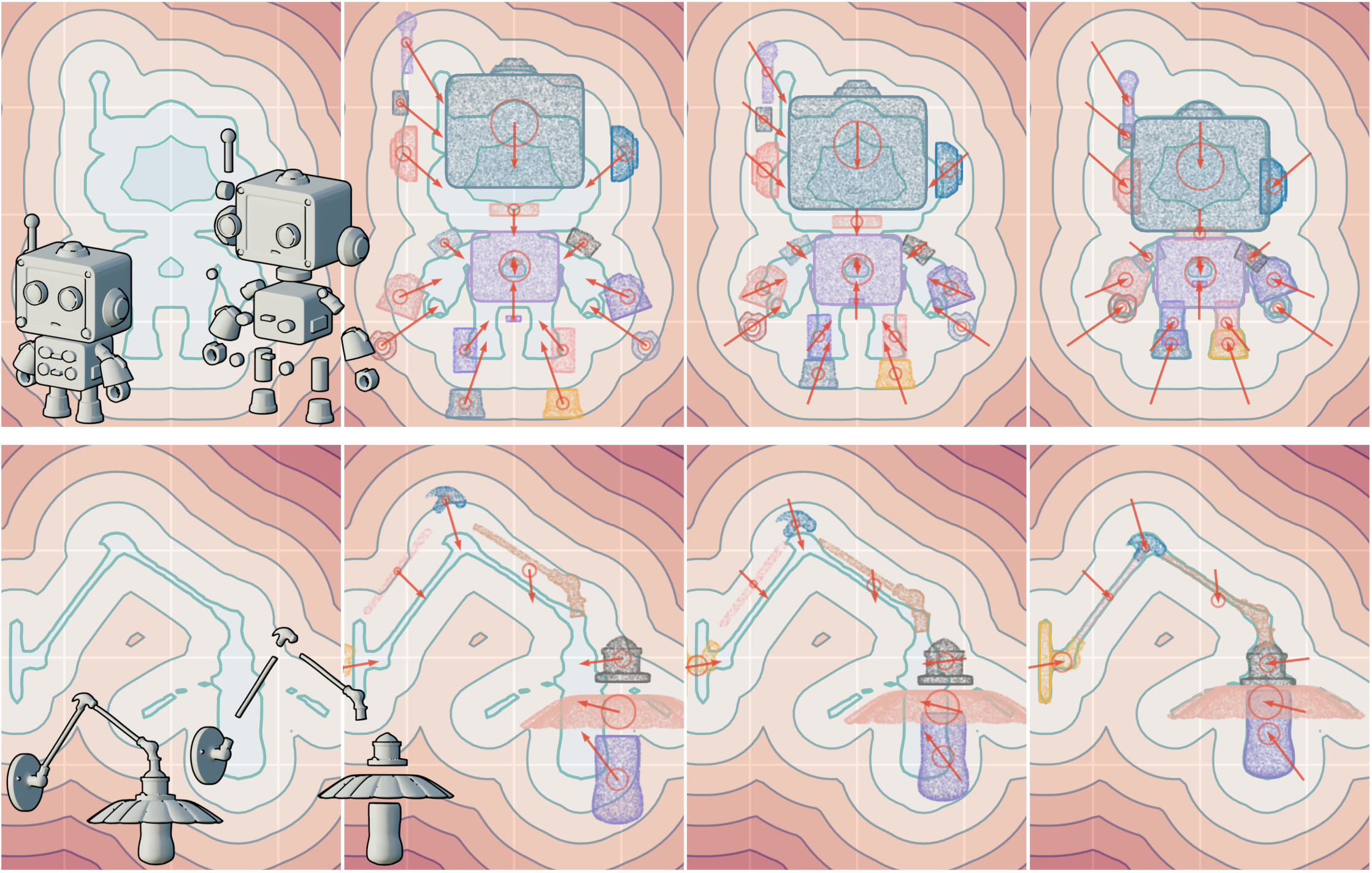}

    \caption{
Visualization of the SDF-based part trajectory tracking process. The figure illustrates this process for two example generated objects, with the cross-section SDF at $t=0$ shown as the background. Each row, from left to right, represents: (1) the assembled and exploded geometry; (2) the parts at 
$t=1$, indicating the fully exploded state; (3) an intermediate state at 
$t=0.5$, showing the parts moving along their trajectories; and (4) the final assembled result at $t=0$. Red arrows denote the optimized trajectories, and part centers are highlighted with red circles. Parts far from the cross-section plane are omitted for clarity.
    }
    \label{fig:trajectory}
\end{figure}

\paragraph{SDF-based Trajectory Optimization.} 
Our approach adopts the SDF representation for geometry generation, and hence accommodates a companion SDF-based part tracking scheme. This volumetric perspective ensures that each part’s position can be optimized to align with its designated region in the final shape, accommodating intentional intersections where necessary.
Specifically, given a generated exploded dynamics $\{\mathcal{M}_{t=0},\dots, \mathcal{M}_{t=1}\}$, we identify all individual parts $\{\bm{P}_i\}$ from the fully exploded state $\mathcal{M}_{t=1}$ through connected component analysis. 
To formalize how each part $\bm{P}_i$ moves from its position $\bm{p}_i^1$ in $\mathcal{M}_{t=1}$ (exploded state) back to $\bm{p}_i^0$ in $\mathcal{M}_{t=0}$ (assembled state), we format a linear parametrization of translation 
$\bm{p}_i^t = \bm{p}_i^0 + \bm{v}_i (1-t)$,
where $\bm{v}_i$ is the translation vector of $\bm{P}_i$ as $t$ goes from $1$ to $0$. Notice this parametrization is valid as our training data assumes linear translation of structural parts.
Here, our target is to optimize a per-part translation vector $\bm{v}_i$. Notably, the SDF values serve as a natural metric for evaluating the fitness of each part, since well-aligned parts exhibit SDF values near zero at the boundaries. Thus, we randomly sample a surface point cloud $ \tilde{\bm{P}_i} $ from $ \bm{P}_i $ and minimize the absolute SDF value on the motion path of the point cloud across frames. %
The optimization of $\bm{v}_i$ is formulated as: 
\begin{equation}
    \{ \bm{v}_i \} \leftarrow  \arg \min  \sum_{t} \sum_{i} \big | \text{QuerySDF}(\mathcal{M}_t,\tilde{\bm{P}_i} + \bm{v}_i (1-t)) \big |  ,
\end{equation}
where $\text{QuerySDF}(\mathcal{M}_t,\cdot)$ represents querying SDF values from the corresponding 3D points of the watertight mesh $\mathcal{M}_t$.
Fig.~\ref{fig:trajectory} illustrates examples of optimized trajectory with target SDF.

\paragraph{Stop Overlapped Point Gradients.} 

The SDF-guided optimization works effectively when there is no overlap between parts. However, when two parts $\bm{P}_i$ and $\bm{P}_j$ overlap, the surface points within their intersection region (identified by negative SDF values) become invalid for optimization. Only the ``frontier'' points on the actual boundaries provide meaningful gradient signals. Hence, we mask out any surface points located inside another part (i.e., those with negative SDF values) during the loss computation, focusing the optimization on the boundaries. This results in better tracking accuracy, which will be evaluated in Sec.~\ref{sec:evaluation}.
The optimized translation vectors of each part ensure a plausible reassembly or disassembly path, so we can generate intuitive exploded and reassembled trajectories that maintain the structural coherence of the original asset.

\section{Controllable Generation}
\label{sec:control}

In creative workflows, artists often require a high degree of control of how an object decomposes. To meet this need, we introduce two complementary schemes to enable our generative explode dynamics for controllable generation via spatial prompts, including bounding boxes and surface regions. Moreover, our generative decomposition inherently preserves geometric semantics and spatial relationships in the exploded dynamics dataset. We hence integrate our framework with 2D feature extractors (e.g., DINOv2~\cite{oquab2023dinov2}) and multi-modal models (e.g., GPT-4 family~\cite{achiam2023gpt4}) for further intuitive and seamless creation.

\begin{figure}
    \centering
    \includegraphics[width=1\linewidth]{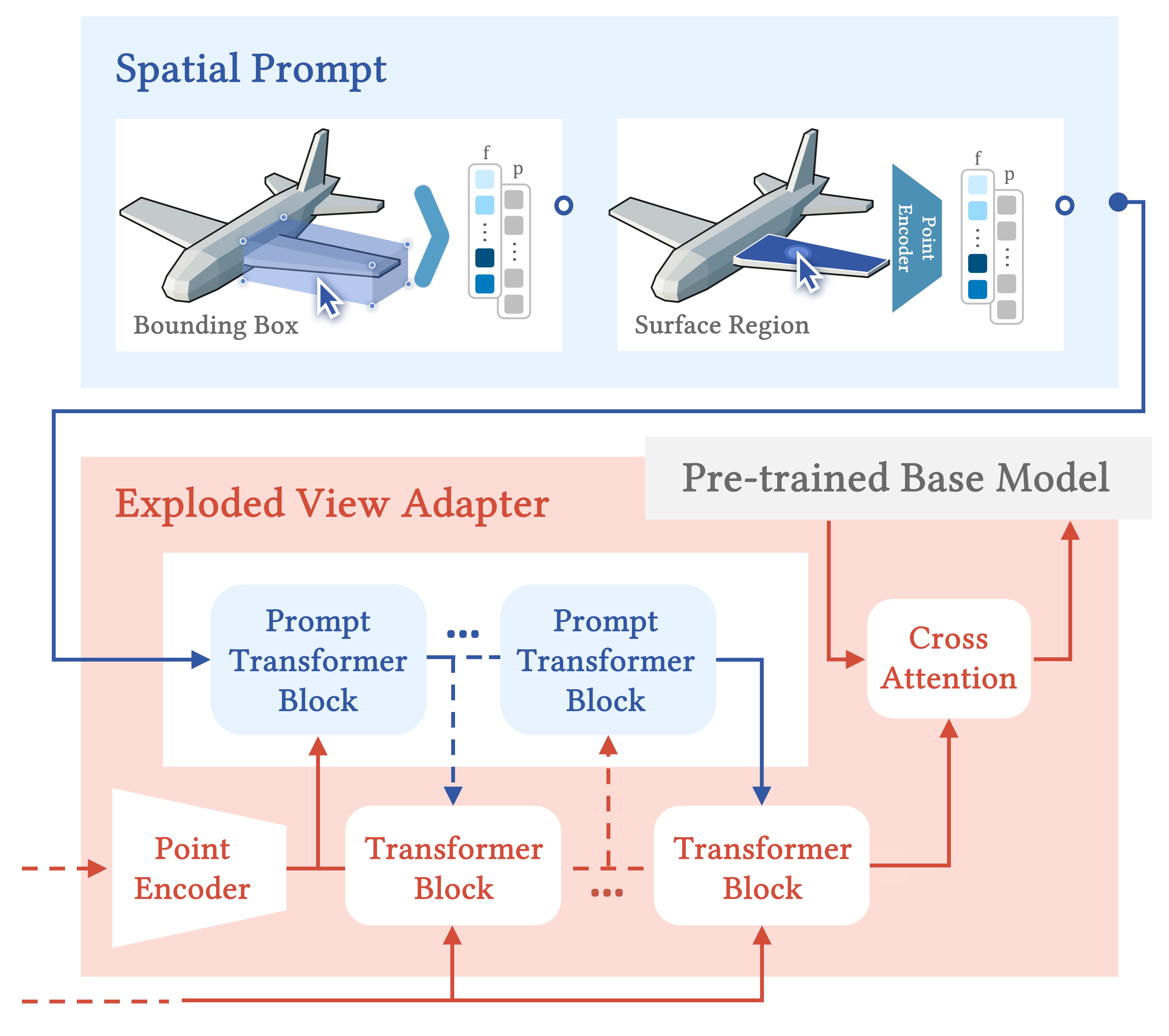}

    \caption{
    The architecture of the extended exploded view adapter for processing spatial conditions. The conditions, provided as either a 3D bounding box or a selected surface region on the input geometry, is first encoded into tokens. These tokens are then fed into the Prompt Transformer Blocks, which exchange information with the Transformer Blocks of the exploded view adapter. This enables the system to dynamically adjust the exploded view generation process based on the user-defined constraints.
    }
    \label{fig:prompt}
\end{figure}

\begin{figure}
    \centering
    \includegraphics[width=1\linewidth]{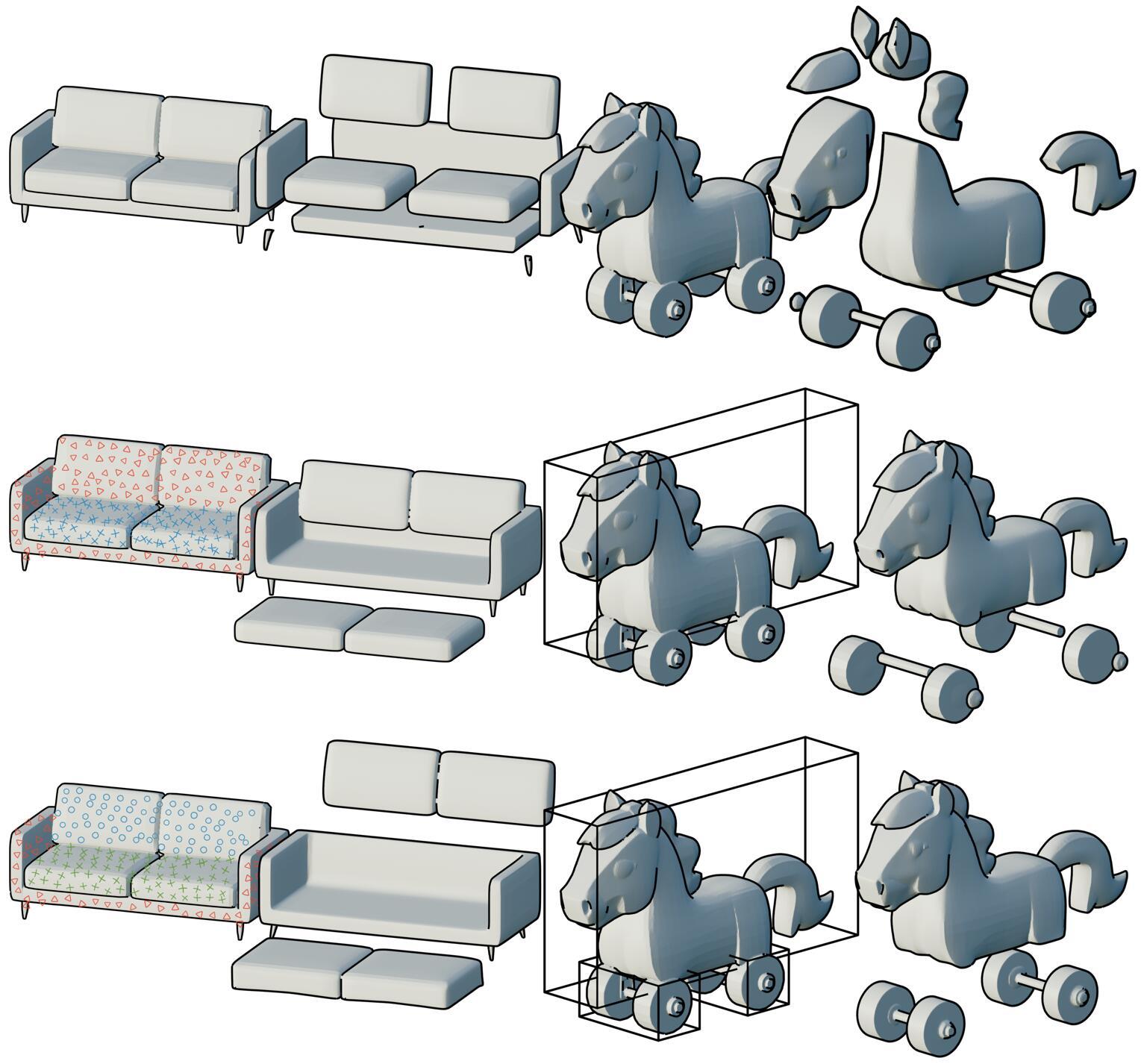}

    \caption{
    Effect of spatial prompt control for exploded view generation. The figure illustrates two distinct examples: a sofa (left) and a wooden horse toy (right). The first row displays the generated exploded views without any spatial prompt input, serving as the baseline. The middle and bottom rows show the effects of varying spatial prompt settings, including different surface regions and varying bounding boxes. These prompts enable users to selectively control which parts to exploded, demonstrating the controllability and flexibility of our method.
    }
    \label{fig:prompt_result}
\end{figure}

\subsection{Spatial Control}

We provide two kinds of natural conditioning prompts for spatial control in our generative exploded dynamics, i.e., 3D bounding boxes and surface regions on the mesh. Bounding boxes allow users to specify a volumetric region, even if the original geometry lacks an internal structure (for instance, if a table with a drawer is only modeled externally, which will be discussed in Sec.~\ref{sec:vis}). Surface regions, on the other hand, provide more precision for cases in which an artist wants to isolate a finely detailed area on the object’s surface.

To support the spatial condition, we use the positions of points as prompts, i.e. the diagonal corners of bounding box and sampled points for surface regions, and extend the exploded view adapter by incorporating a dedicated transformer branch to handle the spatial prompts. 
More specifically, we create new transformer blocks to process the spatial prompts. We apply positional embedding $\text{PosEmb}(\cdot)$ to the bounding box corners as tokens, and use a point encoder (as in Eqn.~\ref{eq:pointencoder}) to encode the surface point-cloud into tokens. 
To differentiate between multiple spatial prompts, we add unique positional embeddings. The encoded tokens are then integrated with the geometry features $\bm{G}$ through interleaved cross-attention mechanisms, allowing the model to effectively interpret and utilize the spatial guidance provided by the user, as shown in Fig.~\ref{fig:prompt}.

During training, both bounding boxes and surface regions are randomly selected from the training data, with varying numbers of prompts per instance to enhance the model’s flexibility. 
Additionally, an auxiliary binary token is included to indicate whether the bounding boxes correspond to all parts to be generated, or the unselected regions should still be exploded automatically.
This training strategy ensures that the model can handle an arbitrary number of spatial prompts during inference, providing users with the ability to control the decomposition process according to their specific requirements. As illustrated in Fig.~\ref{fig:prompt_result}, users can seamlessly guide the generation by specifying spatial regions, resulting in controlled and intuitive exploded views tailored to their creative intentions.
With the spatial conditioning scheme, our exploded generative framework empowers users with the ability to decide how and where an explosion conducts. 
For example, artist can maintain creative oversight by selecting only the wheels of a wooden horse for separation, or merging all body parts into a single chunk. This approach drastically expands the potential use cases and fosters the precise control that designers, hobbyists, and other practitioners often need in real-world scenarios.

\subsection{Cross-modal Creative Framework}
\label{sec:geoencoder}

While bounding-box and surface-region prompts address fundamental controllability requirement, practical workflows often demand more accessible interaction approach. Common users are more comfortable with indicating regions in 2D domain than manipulating 3D space.
Fortunately, our generative model naturally preserves geometric semantics and spatial relationships distilled from a large-sale of 3D data, which can be effectively aligned with 2D feature extractors.
To this end, we can match the specified region rendered in a 2D image to its corresponding location on the 3D mesh, allowing a user to select a part of the object from its rendered view or even a sketch image. 

\begin{figure}
    \centering
    \includegraphics[width=1\linewidth]{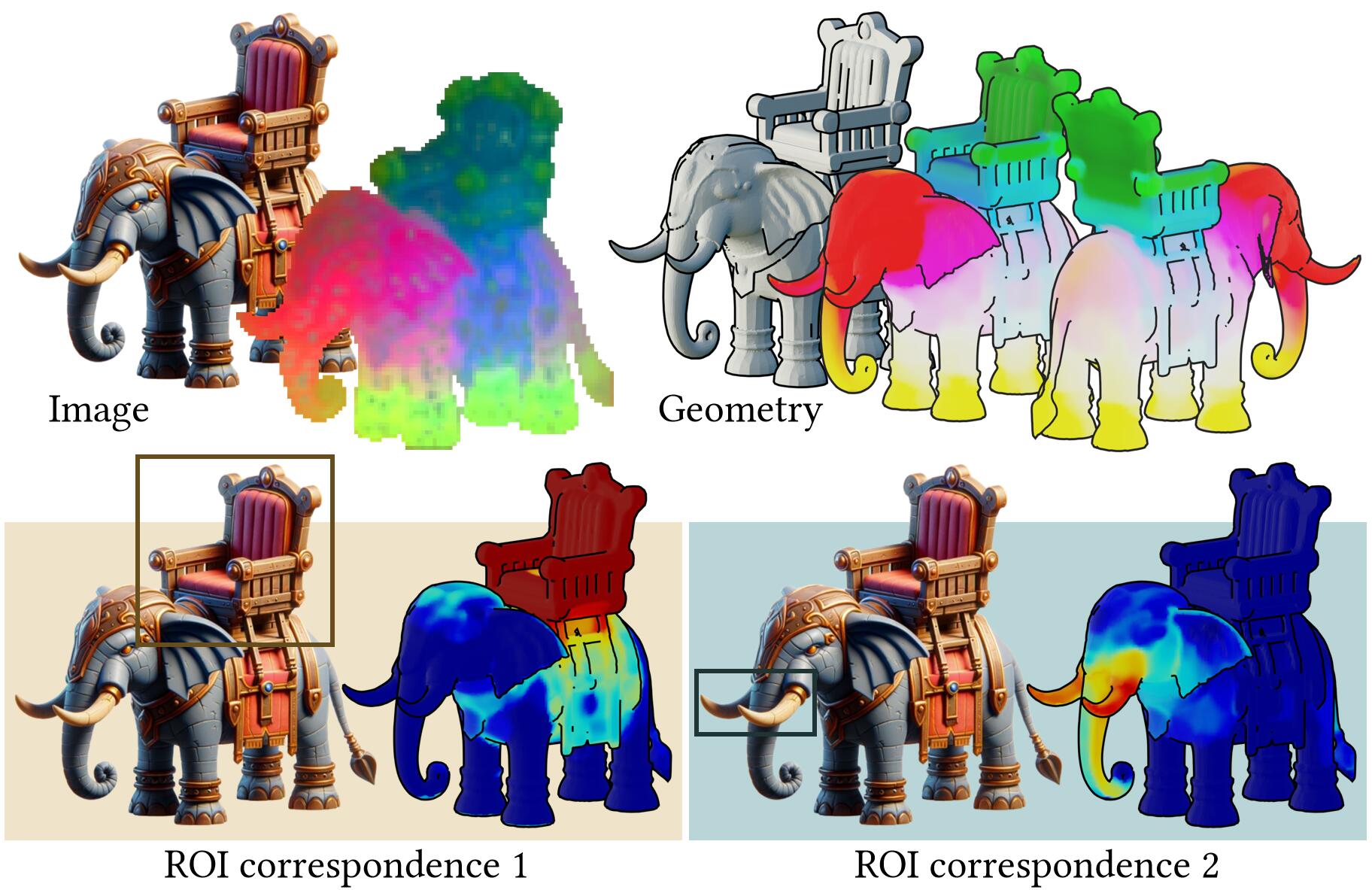}

    \caption{
    2D-3D semantic correspondence is established by aligning features from a 2D image (top left) and its 3D geometry (top right) using geometric feature extractor and DINOv2. The bottom row demonstrates how selected regions of interest (ROIs) in the image corresponds to regions on the 3D geometry.
    }
    \label{fig:dino}
\end{figure}

\paragraph{Geometry and 2D Feature Alignment}
To align 3D geometry with 2D image feature, we re-purpose a VAE decoder $\mathcal{D}$ to produce geometric features aligned with DINOv2~\cite{oquab2023dinov2} rather than produce SDF values.
Formally, for a geometry latent code $\bm{Z}$ and point cloud $\bm{p}$ on the corresponding mesh surface, we compute:
\begin{equation}
\mathcal{D}_\text{feature}(\bm{Z},\bm{p})=\text{CrossAttn}(\text{PosEmb}(\bm{p}),\text{SelfAttn}^{24}(\bm{Z})), 
\end{equation}
which result in feature vectors at surface points $\bm{p}$.
For training $\mathcal{D}_\text{feature}$, we render the corresponding 3D asset into a color image $\bm{I}_\text{RGB}$ and a depth map $\bm{I}_\text{depth}$, extract the DINOv2 features of $\bm{I}_\text{RGB}$, and un-project $\bm{I}_\text{depth}$ back to 3D points $\bm{p}_\text{depth}$. We compute the features of $\bm{p}_\text{depth}$ using $\mathcal{D}_\text{feature}$, match them with the DINOv2 features:
\begin{equation}
\mathcal{L}_\text{align}=\sum \big \| \mathcal{D}_\text{feature}(\bm{Z},\bm{p}_\text{depth})- \text{DINOv2}(\bm{I}_\text{RGB}) \big \|_2.
\end{equation}
Once trained, we can use the rendered images or concept images to specify semantic regions of interest (e.g., via a segmentation tool like SAM2~\cite{ravi2024sam2}), and identify the corresponding 3D regions through feature similarity of sampled surface points with the DINOv2 features in the 2D region. 
This involves selecting 3D features within a certain distance threshold of any 2D ROI features.
As illustrated in Fig.~\ref{fig:dino}, this bridging mechanism enables guidance through intuitive 2D annotations, and hence lowers the usage barrier of controllable exploded dynamics generation.

Building upon the geometry feature alignment strategy, our system can be easily integrated with broader generative pipelines, i.e., a large multi-modal generative model that synthesizes 3D objects from text or images. For example, a creator can first produce a virtual asset of a chair using an image prompt—leveraging high-level attributes such as style, color, or shape, and then feed this newly generated 3D mesh into our generative exploded dynamics. At that point, additional bounding boxes, surface regions, or 2D region-of-interest selections can specify precisely which parts of the chair explode or how the explosion proceeds. This creates an end-to-end workflow where novel 3D objects are designed and interactively decomposed, seamlessly bridging initial concept generation with precise part-level manipulation.
Collectively, these strategies highlight the value of controllability in exploded dynamics generation. By enabling user interaction through bounding-box prompts, surface regions, or 2D region selections, our approach seamlessly integrates advanced part-level decomposition into existing creative workflows, enabling intuitive and efficient exploration of design arts, manufacturing, or educational tasks.

\begin{figure}
    \centering
    \includegraphics[trim={100 0 50 0},clip,width=1\linewidth]{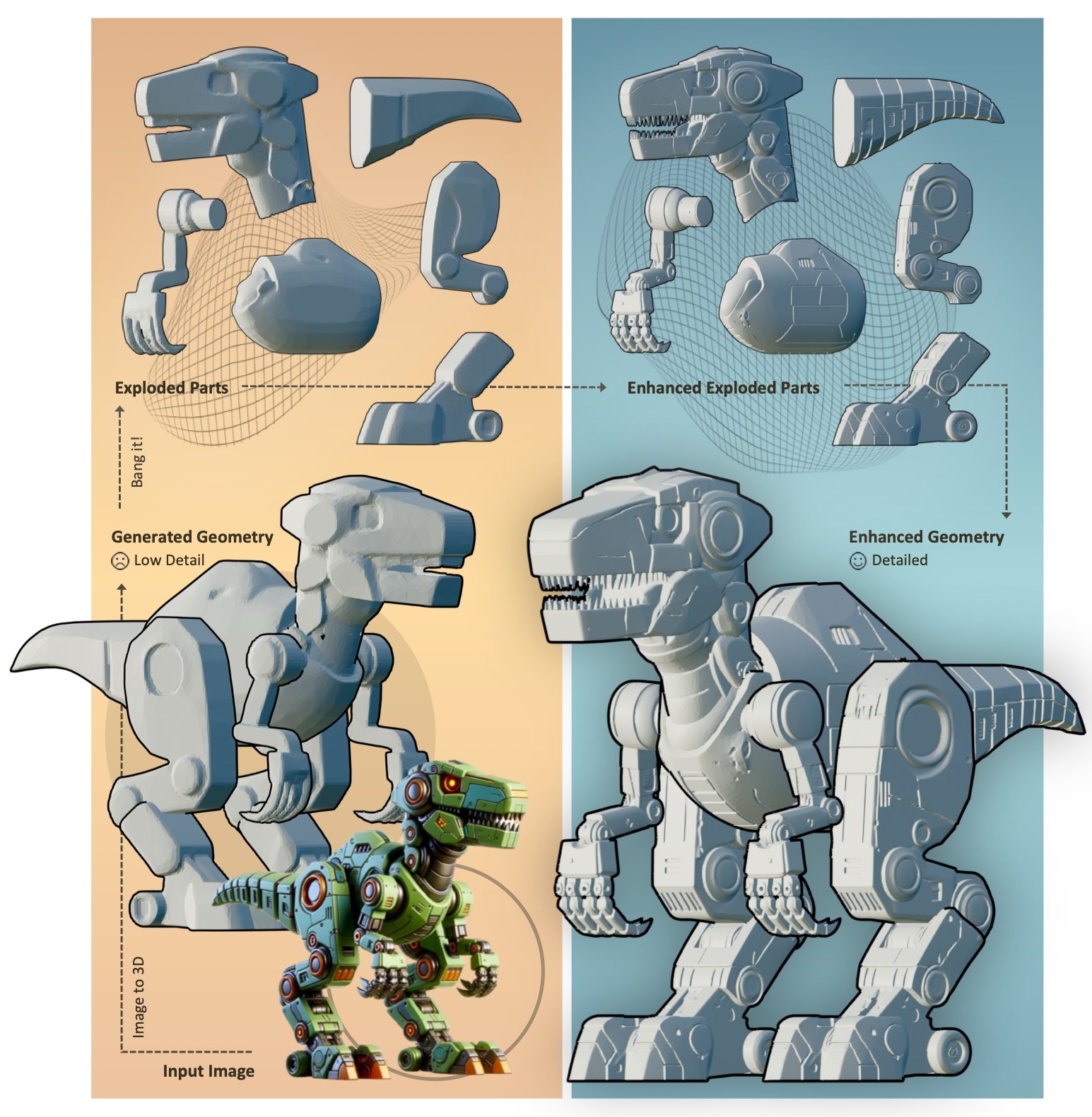}

    \caption{
    Enhanced geometry quality through per-part enhancement. We start with an input image of a robotic dinosaur from TRELLIS~\cite{xiang2024trellis} (bottom left). We generate a 3D geometry (center left) and explode it into parts (top left). Each part is then regenerated based on its coarse geometry, enhancing its detail (top right). Finally, the enhanced parts are reassembled, resulting in a more detailed and accurate 3D geometry (center right) that closely matches the input image.
    }
    \label{fig:part_regen}
\end{figure}

\section{Applications}
The unique capacity of BANG to transform complex 3D assets into detailed, interpretable parts benefits various fields, from industrial design to virtual reality and digital art. By enabling granular part-level control, smooth temporal transitions, and implicit structural awareness, BANG with generative exploded dynamics empowers users to efficiently manipulate and interpret complex 3D assets. 
In the following sections, we illustrate three key applications of BANG, showcasing its impact on component-driven 3D creation, understanding, and manufacturing workflows. It can significantly streamline creations, and enhance collaborative design and immersive experiences, highlighting its huge potential to drive long-term innovation in 3D creation and interaction.

\subsection{Per-part Geometric Detail Enhancement}
\label{sec:partregen}

Our framework facilitates a full cycle of part disassembly, per-part refinement, and subsequent reassembly for geometric detail enhancement. The reliance on a signed distance function (SDF) representation within a normalized space of $[-1, 1]^3$ introduces challenges in simultaneously modeling the entire structure and capturing intricate surface details. By isolating each component in its exploded state, our method re-scales individual parts to the normalized space and reconditions them based on coarse geometry and corresponding image regions, thus enabling high-fidelity local refinements.
During this refinement stage, defects can be corrected while local geometries are enhanced, and each part is then reassembled through the trajectory optimization process in Sec~\ref{sec:tracking}. As a result, the regenerated components align seamlessly with the original global structure, producing a multi-part object that preserves fine-grained detail across all regions.

We illustrate this process in Fig.~\ref{fig:part_regen}. Given an input geometry generated by our base model, our approach first explodes it into individual parts. Each part is then scaled into the normalized space and regenerated based on its coarse geometry and corresponding image regions, producing highly detailed geometry. Finally, the enhanced parts are re-assembled into their original position. This approach achieves a higher level of detail compared to those single-mesh pipelines that generate the entire geometry as a whole, and further facilitates artist-friendly topologies and enables part-specific animations. By focusing on part-level regeneration, our method enhances both the visual quality and functional versatility of 3D assets, surpassing previous methods that are limited by resolution and single-mesh representations.

\begin{figure}
    \centering
    \includegraphics[width=1\linewidth]{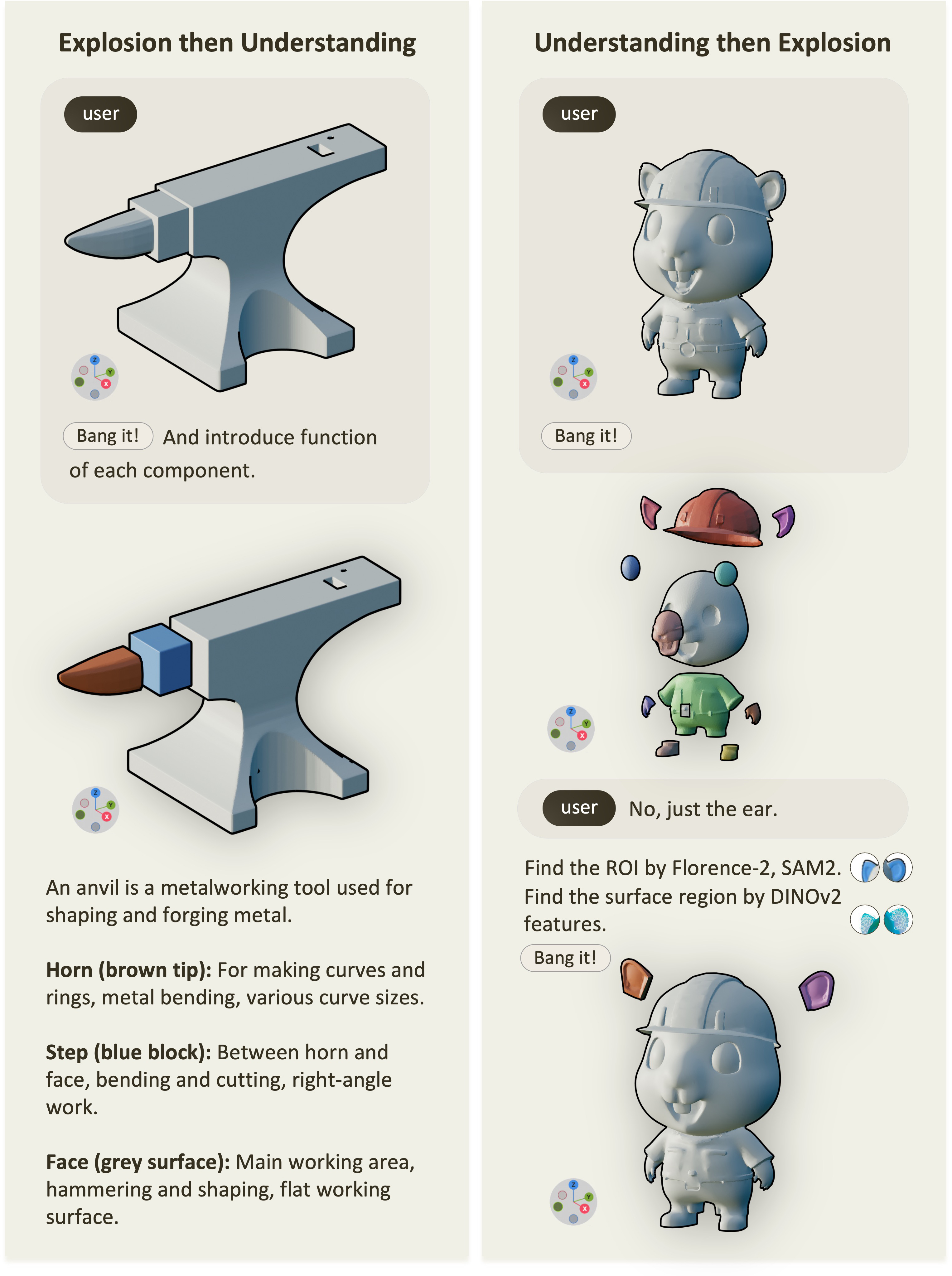}

    \caption{
    Interactive exploded views via chatbot integration: Our framework combines object understanding and generative explosion through interactive dialogue with a Chatbot. We showcase two interaction paradigms: ``Exploded then Understanding'' (left), where an automatic explosion generates functional descriptions, and ``Understanding then Explosion'' (right), where user queries guide the decomposition of specific parts.
    }
    \label{fig:chat}
\end{figure}

\begin{figure}
    \centering
    \includegraphics[width=1\linewidth]{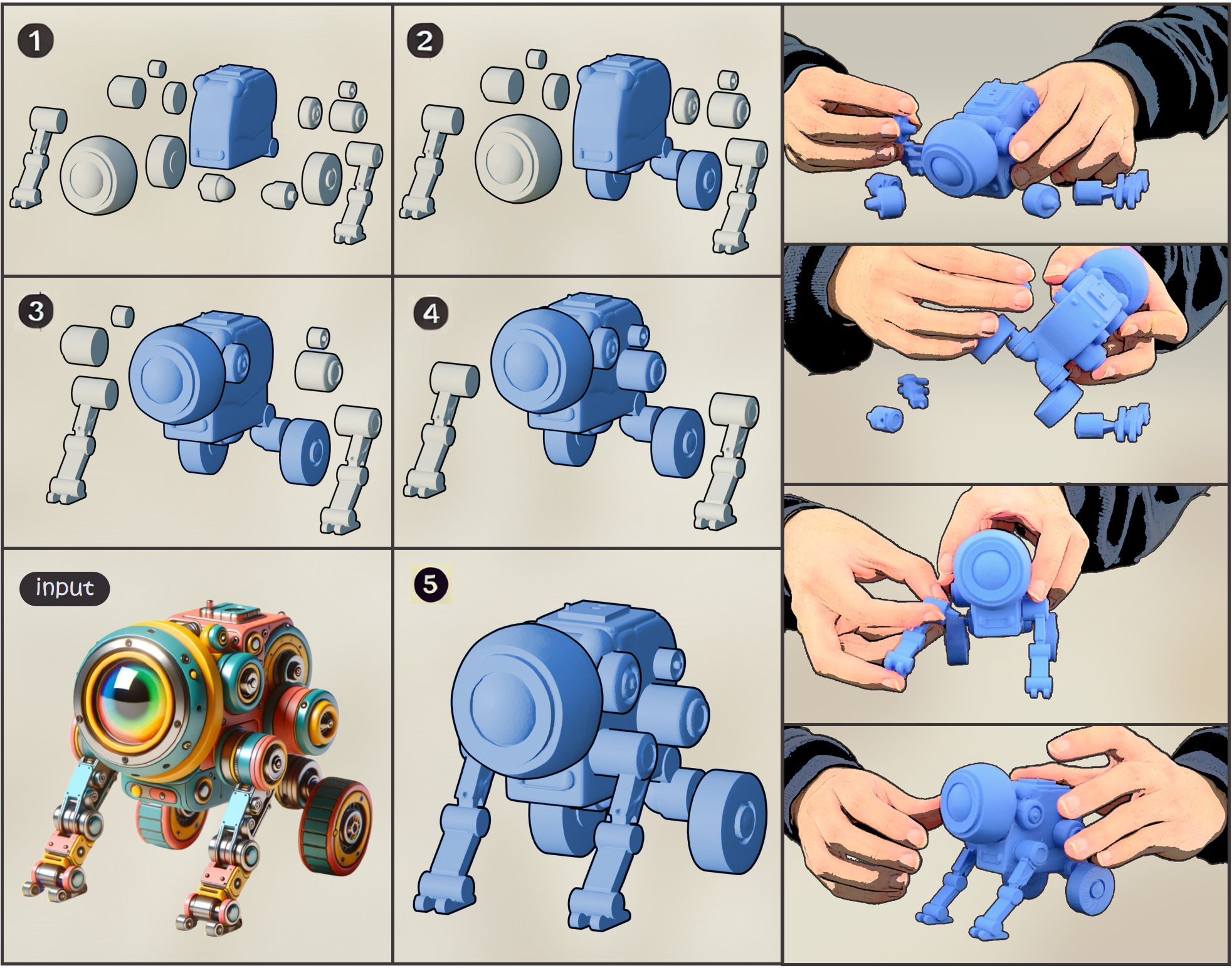}

    \caption{
Expedite physical prototyping of combinable structures. Each part of a robot, generated from a cute robot design (featured in TRELLIS~\cite{xiang2024trellis}), is 3D printed using the X1 Carbon~\cite{bambu_lab_x1_carbon} and then assembled (right column). The interlocking structures between parts is programmatically generated, allowing parts to be seamlessly connected and assembled after printing. This demonstrates our approach’s ability to preserve structural integrity while enabling easy post-printing assembly.
    }
    \label{fig:3dprinting}
\end{figure}

\begin{figure*}
    \centering
    \includegraphics[width=1\linewidth]{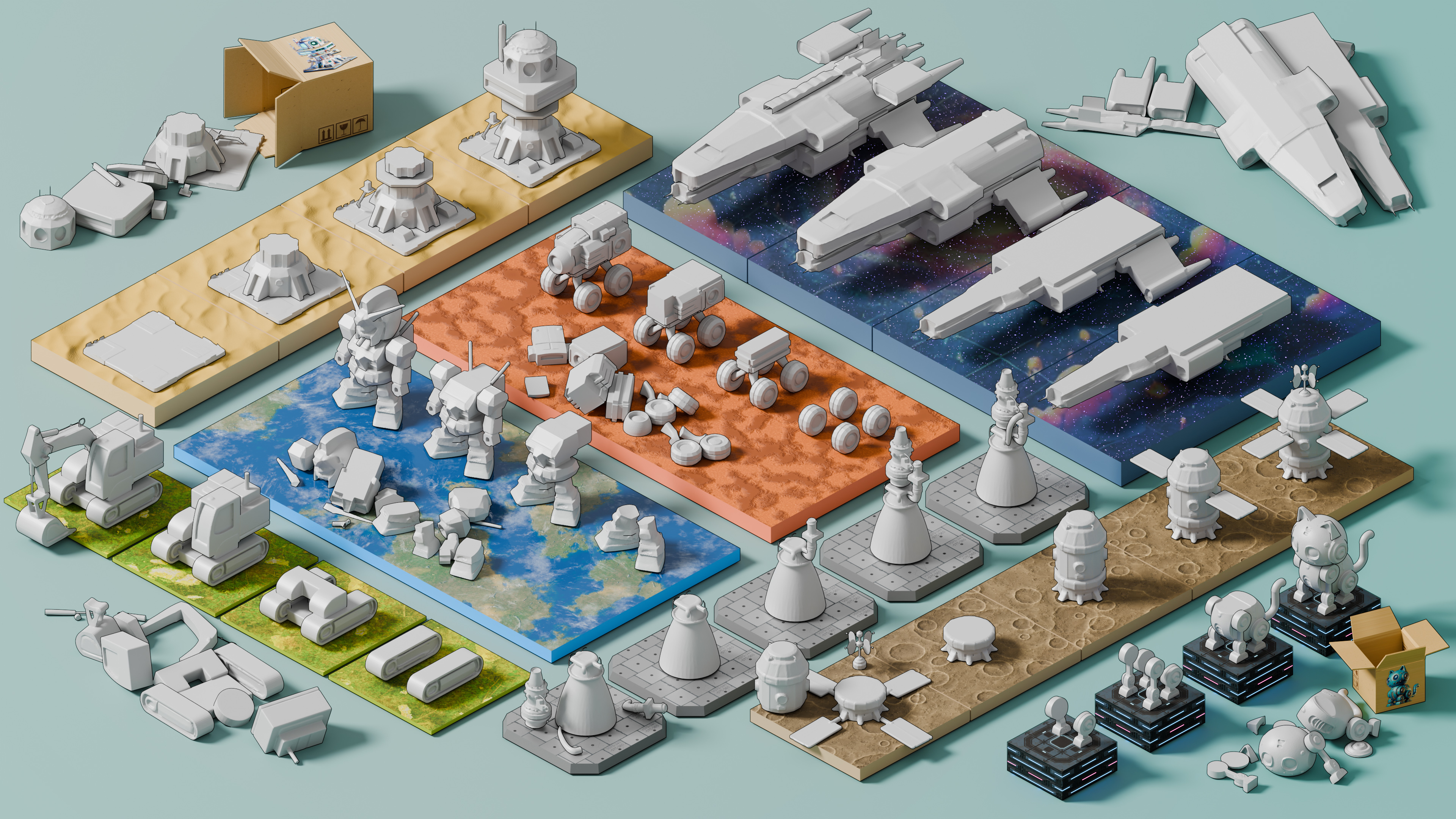}

    \caption{
A fictional journey from Earth's surface to the far reaches of space, celebrating humanity's boundless ingenuity and spirit of discovery. Each object is generated from a concept image and illustrated in four assembly states, using parts generated from our \textit{Generative Exploded Dynamics}.
}
    \label{fig:ga_trajectory}
\end{figure*}

\begin{figure*}
    \centering
    \includegraphics[width=1\linewidth]{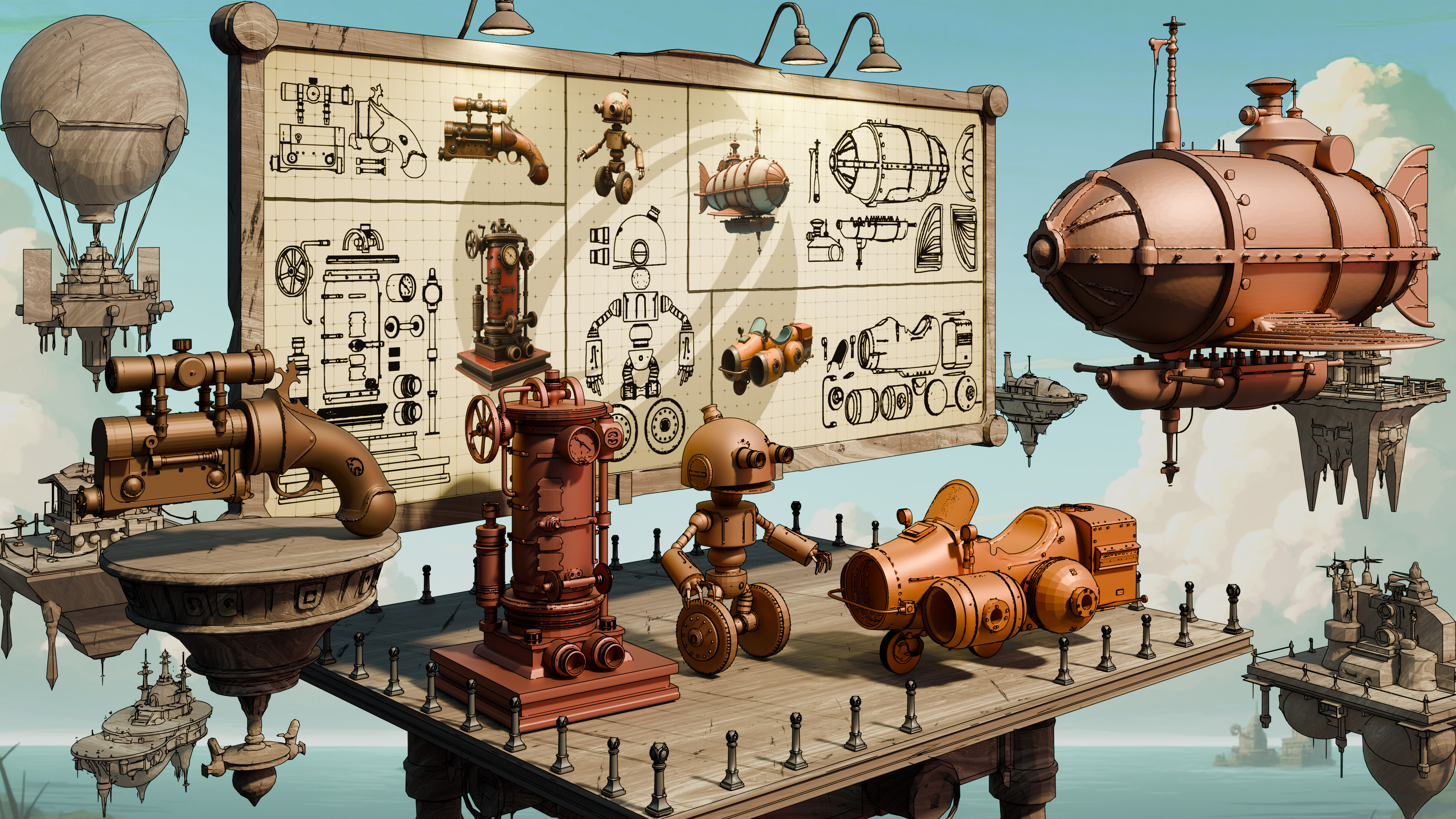}

    \caption{
A steampunk workshop, where blueprints transform into tangible reality, powered by our \textit{BANG} framework. Each asset begins as a concept image generated by FLUX~\cite{flux2023}, is then transformed into an integral 3D mesh via our base generative model, and subsequently exploded into parts and are meticulously enhanced part-wise for maximum visual fidelity. The generated exploded structures are displayed against the backdrop, showcasing the enhanced details achieved through our exploded-enhance pipeline.
    }
    \label{fig:ga2}
\end{figure*}

\subsection{Multi-modal Integration for Structural Understanding and Control}

The integration of BANG with multimodal large language models (MLLMs) greatly enhances the part-level understanding of 3D objects, bridging the gap between generative 3D creation and semantic comprehension. By leveraging MLLMs such as GPT-4 family~\cite{achiam2023gpt4}, BANG can automatically assign descriptive labels, functional attributes, and contextual information to individual sub-components of 3D meshes, offering a deeper understanding of the object's structure and purpose.
As illustrated in Fig.~\ref{fig:chat}, our framework supports two key interaction paradigms to exemplify how MLLMs can be used in conjunction with 3D geometry.
    \paragraph{Explosion then Understanding} 
    In this paradigm, the 3D object is first decomposed into its constituent parts through the generative exploding process. After the explosion, the system provides detailed textual descriptions and contextual insights for each part, facilitating iterative design and evaluation. The exploded view with clear part decomposition is rendered into images, which are then analyzed by the MLLM to generate these descriptions.
    To ensure clear part identification, each part is assigned a distinct visual marker during rendering, such as color coding or numbered overlays. These annotated images are then provided to GPT-4, enabling it to reference specific parts unambiguously and generate corresponding descriptions, functions, or semantic roles for each.
    \paragraph{Understanding then Explosion} 
    In this paradigm, users interact with the system through natural language commands, guiding the decomposition process based on the object's functional descriptions or relationships between parts. For example, users can specify which parts to isolate or modify, enabling more precise and targeted manipulations. This interaction is facilitated by MLLMs generating text-based instructions, which are then used in combination with models like Florence-2~\cite{xiao2024florence2} for 2D region-of-interest (ROI) selection, SAM2~\cite{ravi2024sam2} for segmentation, and DINOv2~\cite{oquab2023dinov2} with our geometric feature extractor (Sec.~\ref{sec:geoencoder}) to map these selections accurately to the 3D geometry. Spatial prompts are then applied for controllable generation based on these selections.

This multi-modal integration enriches the semantic annotations of 3D objects, providing users with intuitive, flexible control over part-level manipulations.
By linking textual and visual understanding to 3D geometry, our framework opens new possibilities for creative and industrial workflows, enhancing design, analysis, and modeling processes. This fusion of generative and semantic reasoning not only streamlines the development cycle but also fosters more dynamic and collaborative environments, pushing the boundaries of interactive 3D modeling and intelligent system integration.

\subsection{Expedite Combinable Structure 3D Printing}

3D physical prototyping of combinable structures is widely used in industrial design, customizable product development, robotics, etc. It necessitates the segmentation of designs into print-friendly components while ensuring that the final assembly maintains consistency and functionality. BANG inherently supports this requirement by generating part-level meshes with clear separations and precise alignments. These exploded parts can be individually 3D printed, allowing for optimized orientations and tailored material choices for each component. This workflow effectively reduces the need for support materials, mitigates overhang-related printing challenges, and provides flexibility in selecting distinct materials or colors for different parts.
Furthermore, our framework enables the integration of movable joints between components, facilitating dynamic assemblies that can articulate or adjust post-printing. By incorporating such joints, the printed object not only adheres to the intended static design but also gains mechanical versatility, allowing for interactive or functional customization. This enhancement leverages the inherent part structure to achieve both mechanical functionality and aesthetic diversity, thereby increasing the overall utility and appeal of the prototype. As illustrated in Fig.~\ref{fig:3dprinting}, using our approach, one can generate printable parts of a complex toy robot from a single image input, to enjoy hands-on assembly and creation.

\begin{figure*}
    \centering
    \newcommand{\ct}{0.5}
    \begin{overpic}
    [width=1\linewidth]{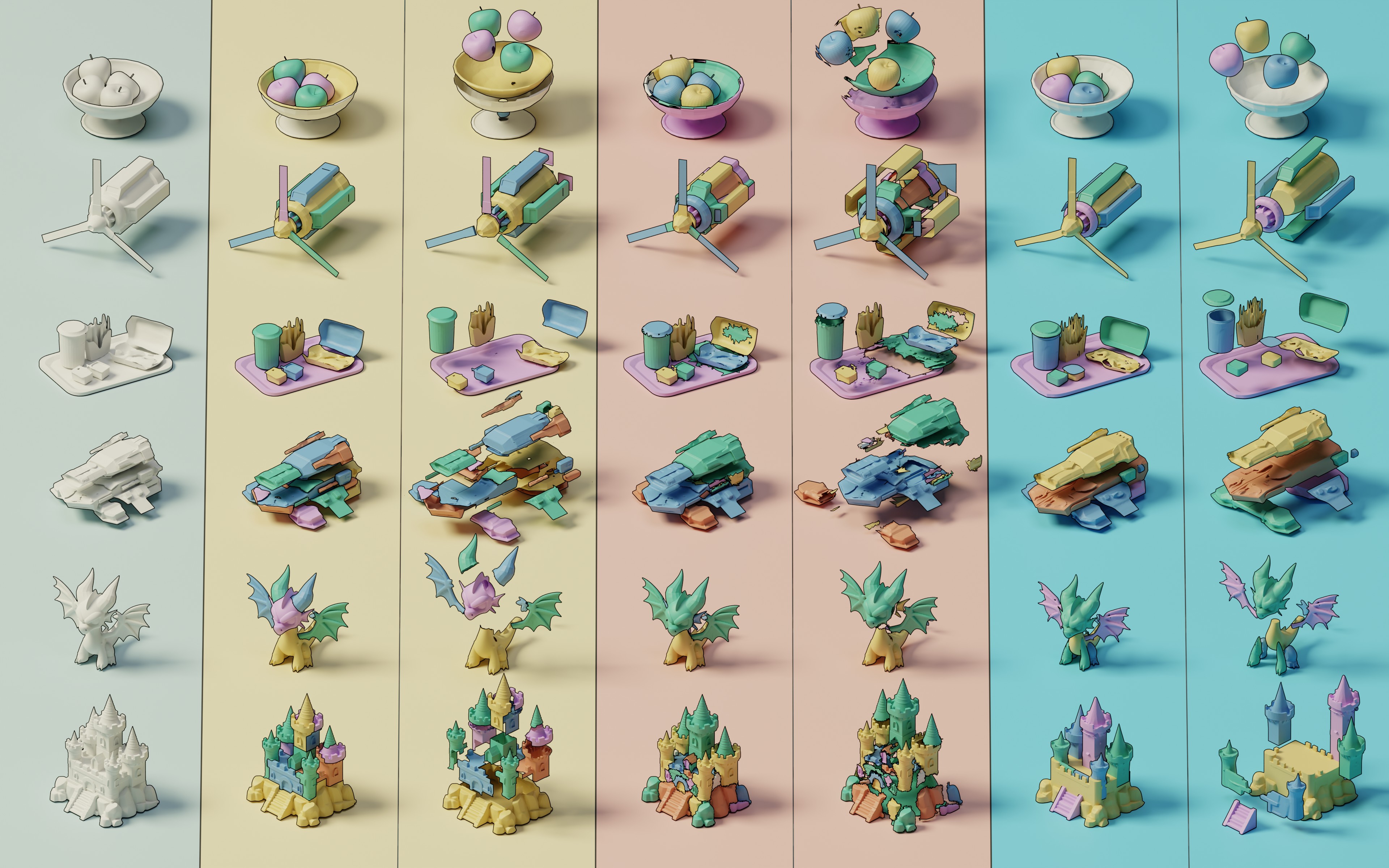}
    \put( 4, \ct){\scriptsize Input geometry}
    \put(16.0, \ct){\scriptsize SAMesh segmentation}
    \put(30.0, \ct){\scriptsize SAMesh exploded view}
    \put(43.9, \ct){\scriptsize SAMPart3D segmentation}
    \put(57.5, \ct){\scriptsize SAMPart3D exploded view}
    \put(74.0, \ct){\scriptsize Ours assembled parts}
    \put(88.0, \ct){\scriptsize Ours exploded view}
    \end{overpic}

    \caption{
Qualitative comparison with part segmentation methods. We compare our approach with part segmentation techniques, including SAMesh~\cite{tang2024segmentanymesh} and SAMPart3D~\cite{yang2024sampart3d}, by displaying both segmentation results and their visualization in exploded views (note these methods are not designed to generate individual parts, we manually separate the segmented components for illustration). For our method, we show the parts both in their assembled state and exploded view, with each part assigned a distinct color.
The test cases include various configurations, i.e.: (1) two intact meshes from PartObjaverse-Tiny~\cite{yang2024sampart3d} (top two rows), featuring clean topology with an artistic style; (2) two re-meshed meshes from PartObjaverse (middle two rows), featuring uniform triangular faces; and (3) watertight meshes generated from concept images using our base generative model (bottom two rows), representing unseen data.
Our method generates part geometries with meaningful decomposition, while the segmentation methods merely separate face regions, failing to preserve the volumetric integrity of individual parts.
    }
    \label{fig:compare}
\end{figure*}

\section{Experiments}

\subsection{Experimental Setup and Implementation Details}
\label{sec:details}

We train our BANG model in a progressive manner, with the first step of pretraining our base generative model following the previous practice~\cite{zhang20233dshape2vecset,zhang2024clay}. Specifically, we train the model on an Objaverse subset~\cite{deitke2023objaverse}, containing $\sim 500,000$ diverse 3D geometries which are processed into a water-tight format. The base model employs a geometry variational autoencoder (VAE) to encode dense point clouds into latent representations of size $2048 \times 64$. 
The VAE encoder consists of 1 cross-attention layer, and the decoder has 24 self-attention layers with 1 cross-attention layer. Both the encoder and decoder use a feature dimension of 512.
The latent diffusion model uses a 24-layer transformer with a hidden size of 2560 and 20 attention heads per layer, following a pre-norm configuration with sequential self-attention, cross-attention, and feed-forward blocks. Each feed-forward block contains an expansion ratio of 4, with GELU activation. The attention layers incorporate qk-normalization, and no gating mechanisms are used. To enhance convergence, we adopt a multi-resolution training schedule, where the latent code resolution is gradually increased from 512 to 2048 during training. Text, image, and point cloud conditioning are handled by CLIP, DINOv2, and a point encoder, with cross-attention for feature modulation.
The training uses AdamW with a learning rate of $1e-5$ and a batch size of 512, conducted over 1600 epochs on 128 GPUs, yielding a robust model capable of generating diverse 3D geometries.
We then train a specialized exploded view adapter to adapt the base model to generate explode views given a mesh. The exploded view adapter consists of 4 transformer layers with a hidden size of 512 to condition the model on both the input geometries, explosion time index, and the expected number of parts. These conditions are embedded via sinusoidal position encodings and added to the latent input. The adapter modulates the base model using cross-attention after the initial embedding layers.
We collect 20k high-quality exploded-view data as described in Sec.~\ref{sec:dataset}, and randomly sample explosion time $t\sim\text{Unif}(0,1)$ during training. We freeze weights of the pretrained base model, and only train the adapter, using AdamW with a learning rate of $1e-5$ and a batch size of 128, over 3000 epochs on 128 GPUs.
Finally, we freeze the weights of the base model and the exploded view adapter, and train the temporal attention module for smooth exploded dynamics generation with the same settings as the exploded view adapter. The temporal attention layers are applied after each cross-attention layer in the base model with the same dimension, ensuring consistent exploded dynamics generation. We send multiple frames in a sequence, with frames count randomly sampled in $[2,5]$ and each explosion time uniformly sampled $t\sim\text{Unif}(0,1)$, at one time and train the temporal attention module using the same settings as the exploded view adapter.
For extraction of geometric features, we distill from DINOv2-Tiny with a 384-dimensional feature, maintaining the same network structure and training settings with the VAE. 
The entire model is implemented in PyTorch and trained on NVIDIA A800 GPUs, utilizing FP16 mixed precision training for computational efficiency.
During inference, exploded dynamics sequences are sampled with 5 frames, setting $\{t\}=\{0,0.25,0.5,0.75,1\}$, using 50 diffusion steps and a DDPM scheduler. Classifier-free guidance is applied with a guidance scale of 7 to enhance generation quality. Gradient clipping (L2 magnitude 1) and learning rate warmup are applied to stabilize training. Due to GPU memory limitations, sequences are limited to 5 frames during training, which provides a balance between quality and memory efficiency and will be discussed in Sec.~\ref{sec:evaluation}.
This approach ensures the generation of high-quality exploded dynamics, with clear part decomposition and temporal consistency, while maintaining computational efficiency.

\subsection{Visualization of Generated Exploded Dynamics}
\label{sec:vis}

We showcase the power of our method in decomposing a complex object into distinct parts in Fig.~\ref{fig:ga_trajectory}. This exploded view demonstrates how, starting from a concept image, our framework generates 3D assets and then breaks them down into individual components. Each part is distinct, making it ideal for applications that require part-level generation and manipulation.

\begin{figure}
    \centering
    \newcommand{\ct}{-2}
    \begin{overpic}[width=\linewidth]{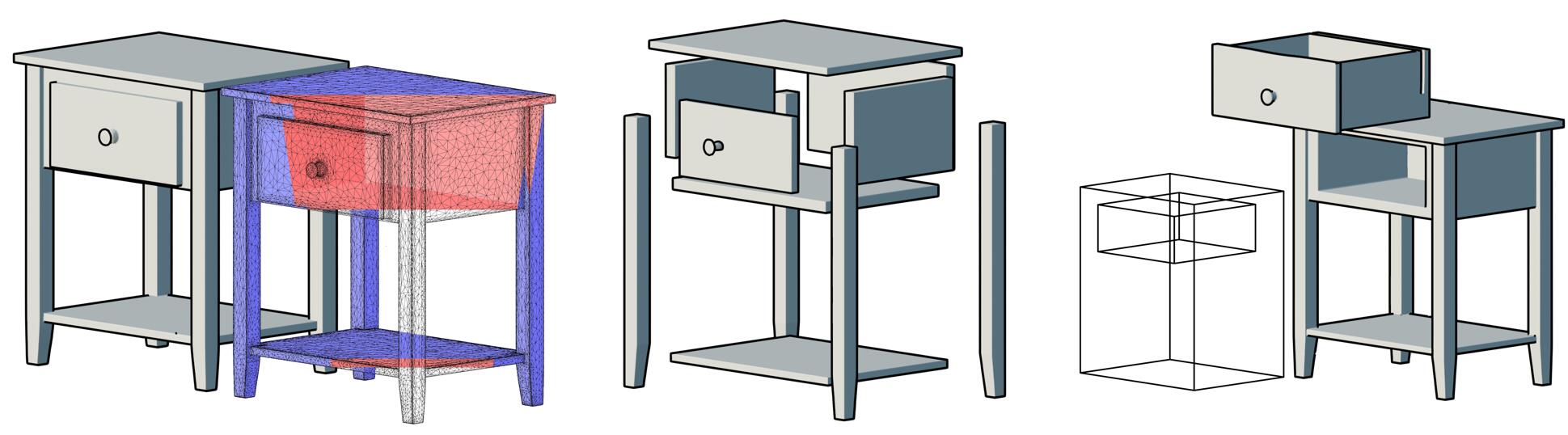}
    \put(5,\ct){\small Input geometry}
    \put(40,\ct){\small Exploded view 1}
    \put(75,\ct){\small Exploded view 2}
    \end{overpic}

    \caption{
    For an input geometry containing only the surface geometry of a table, our approach can generate an exploded view by disassembling the surface mesh into its constituent parts (center). Alternatively, given bounding box prompts, it can infer and generate the corresponding interior structure of the drawer (right).
    }
    \label{fig:drawer}
\end{figure}

Controlling the structure of parts and their positioning in the exploded view is one of the key features of our approach. As shown in Fig.~\ref{fig:prompt_result}, spatial prompts—such as bounding boxes and surface regions—allow users to selectively decompose the object. This enables more targeted control, whether isolating specific parts or choosing how many parts should be exposed. Fig.~\ref{fig:drawer} further illustrates how our system can generate the interior components of an object, such as a drawer, by interpreting user-supplied prompts.

Once exploded, individual parts can be regenerated for higher fidelity. Fig.~\ref{fig:part_regen} shows how we begin with an initial coarse geometry, decompose it, and then regenerate each part for more detailed and accurate surfaces. This approach is exemplified in Fig.~\ref{fig:ga2}, where a steampunk workshop scene showcases how regenerated parts elevate the design’s visual quality. 
Finally, in Fig.~\ref{fig:teaser}, the recursively exploded and regenerated parts come together to form a humanoid mech, demonstrating the practical application of our method in achieving high-quality geometric designs.
This showcases a multi-level creative pipeline: we begin from a concept image, generate a base 3D asset using our pretrained model, apply exploded dynamics with controllable prompts, and recursively enhance and re-explode each part to reveal structural richness, highlighting the iterative generative capabilities of our framework.

Our system also enables interactive exploration, which can deepen understanding through semantic dialogue. Fig.~\ref{fig:dino} illustrates how users can interact with exploded views, gaining a better understanding of the individual components. Additionally, Fig.~\ref{fig:chat} demonstrates the integration of a chatbot, allowing users to request specific information about parts or modify the exploded structure interactively. This interaction bridges the gap between 3D generation and semantic understanding.

Another practical application of our method is in 3D printing. Fig.~\ref{fig:3dprinting} illustrates how exploded parts are generated and printed individually, with optimized orientations and material choices. This process ensures that combinable parts can be assembled easily post-printing while maintaining structural integrity and visual coherence.

\subsection{Structural Segmentation Comparison}
Our BANG is designed for part-aware 3D generation, and there is no exact baseline on this task currently available for comparison. 
To evaluate its effectiveness, we instead compare our method with leading surface segmentation techniques, as part decomposition is a key aspect of our approach.

\begin{figure}
    \centering

    \includegraphics[width=1\linewidth]{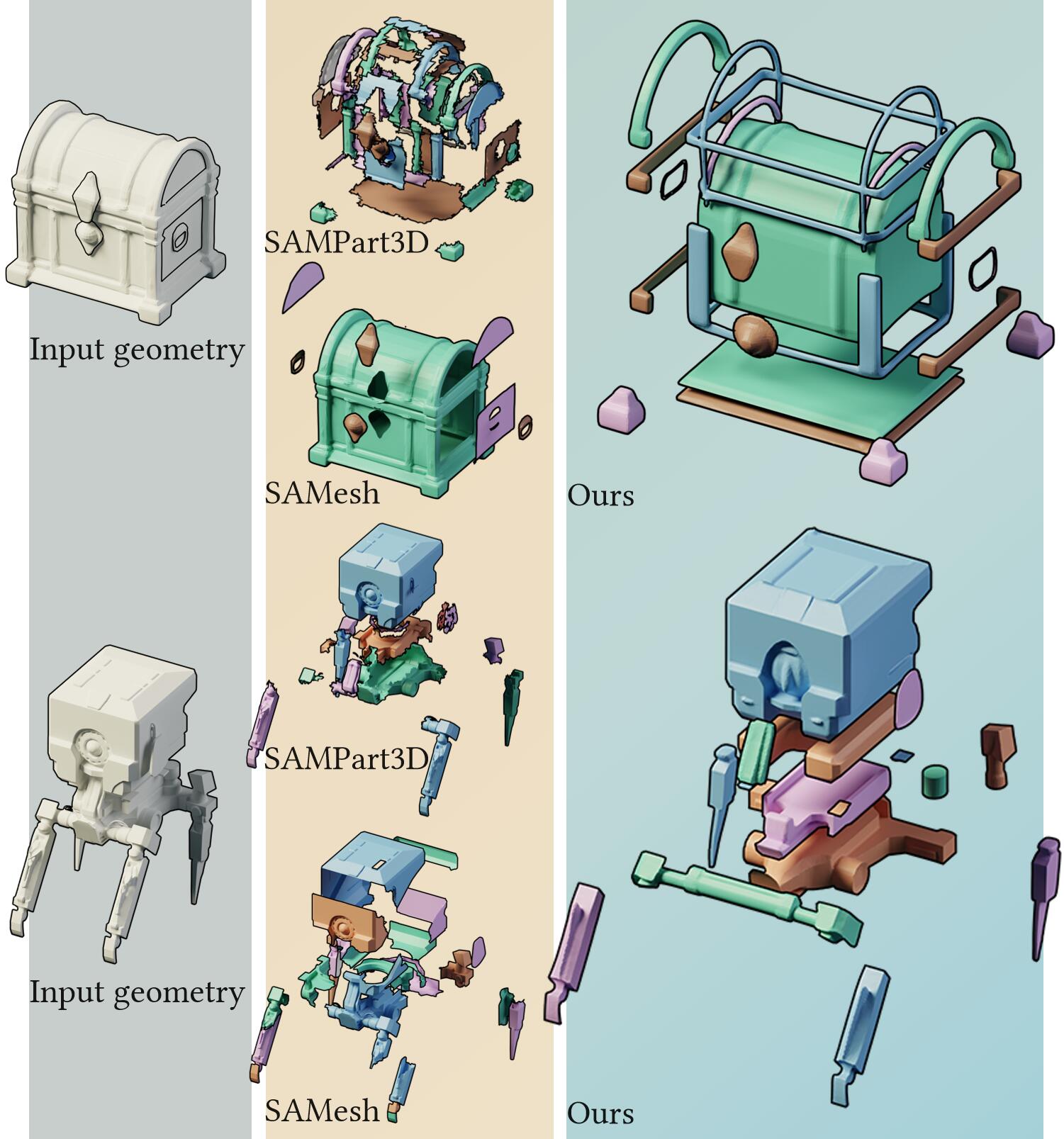}

    \caption{
Comparison on complex generated 3D assets. While segmentation methods exhibit fragmented patches, inconsistent part groupings, and jagged segmentation boundaries, especially under structural complexity, our method produces clean, volumetric part decompositions with consistent semantics and clear structural logic. 
}
    \label{fig:compare_complex}
\end{figure}

We compare BANG with two prominent 3D part segmentation methods: SAMesh~\cite{tang2024segmentanymesh} and SAMPart3D~\cite{yang2024sampart3d}. Fig.~\ref{fig:compare} and Fig.~\ref{fig:compare_complex} showcases these comparisons across different types of input geometries, including meshes from PartObjaverse-Tiny~\cite{yang2024sampart3d} (with artist-crafted topology, not included in our training data), remeshed datasets with uniform triangular faces, and watertight assets generated by our base model.
SAMPart3D is applied to textured assets, while SAMesh and our framework take pure geometry as input. For visualization, the baseline methods display both their segmentation results and manually separated exploded views, whereas our method directly generates exploded views automatically. For both methods, hyperparameters were tuned to yield a segmentation with moderate part granularity.

While SAMesh and SAMPart3D produce reasonable segmentations for simple objects, they struggle with more complex geometries, such as mechanical parts or castle towers. These methods often exhibit inconsistent results due to the limitations of 2D segmentation from multi-view rendered images, and their performance degrades further on non-artist-created triangular meshes. Furthermore, these segmentation methods produce surface-based results—isolating only face regions without any volume or interior structures, limiting their applicability for tasks requiring volumetric part representation.
In contrast, BANG consistently produces high-quality part decompositions across all test cases, maintaining robust part-level generation and volumetric understanding throughout.

\paragraph{User Study}
To further assess the effectiveness of our method, we conducted a user study where 50 participants were shown result produced by BANG, SAMesh, and SAMPart3D on 10 generated assets, and asked to evaluate which segmentation method best aligned with intuitive part decomposition and offered superior visual appeal. 
Notably, our method achieved this with significantly lower computational cost, averaging 45 seconds per asset, compared to 386 seconds for SAMesh and 940 seconds for SAMPart3D.
The results demonstrated a clear preference for our method, with $65.5\%$ of users favoring BANG's generated exploded views. $26.2\%$ of users selected SAMesh, which benefits from multi-view segmentation and classical mesh face processing techniques like smoothing, splitting and graph-cut, offering smooth transitions between parts. $8.3\%$ of users preferred SAMPart3D, which is based on a per-asset MLP learning that is resource-intensive and produces segmented outputs with more noise at the part boundaries. These results highlight that, while SAMesh and SAMPart3D provide reasonable part segmentation for simple geometries, BANG excels in producing more consistent, intuitive, and aesthetically pleasing part decompositions across a wider range of 3D assets.

\subsection{Evaluations}
\label{sec:evaluation}

To quantitatively assess the quality of the generated exploded dynamics sequences and facilitate systematic comparisons, we establish a comprehensive evaluation framework with carefully designed metrics. We select 50 objects from the PartObjaverse-Tiny~\cite{yang2024sampart3d} dataset, which were not included in our training data, as the evaluation set. 
Each asset in PartObjaverse-Tiny contains high quality human-annotated parts.
For each object, we generate exploded view sequences conditioned on ground truth bounding boxes and evaluate the performance using part trajectory tracking. Specifically, we assess three key metrics after transforming the parts back to their original $t=0$ positions: generation time cost, weighted IoU (wIoU), and SDF objective.
We define wIoU as the weighted intersection-over-union between predicted and ground truth bounding boxes for each part. The formula is given by:
\begin{equation}
    \text{wIoU} = \sum_{i} \frac{V_i \cdot \text{IoU}(\bm{B}_i, \bm{B}_i^{\text{gt}})}{\sum_{j} V_j}
\end{equation}
where $V_i$ represents the convex hull volume of the $i$-th part, $\bm{B}_i$ and $\bm{B}_i^{\text{gt}}$ are the predicted and ground truth bounding boxes, respectively. This metric quantifies the accuracy of part localization after explosion.
The SDF objective is introduced to evaluate the geometric alignment between the fitted and actual surfaces of the parts:
\begin{equation}
    \text{SDF}_{\text{obj}} = \frac{1}{|\mathcal{P}|}  |\text{QuerySDF}(\mathcal{M},\mathcal{P})|
\end{equation}
where $\mathcal{P}$ is the set of sampled points on the fitted surface, and $\text{QuerySDF}(\mathcal{M},\cdot)$ calculates the signed distance to the ground truth surface. This objective quantifies how closely the generated parts align with the true surface geometry.

\begin{figure}
    \centering
    \includegraphics[width=1\linewidth]{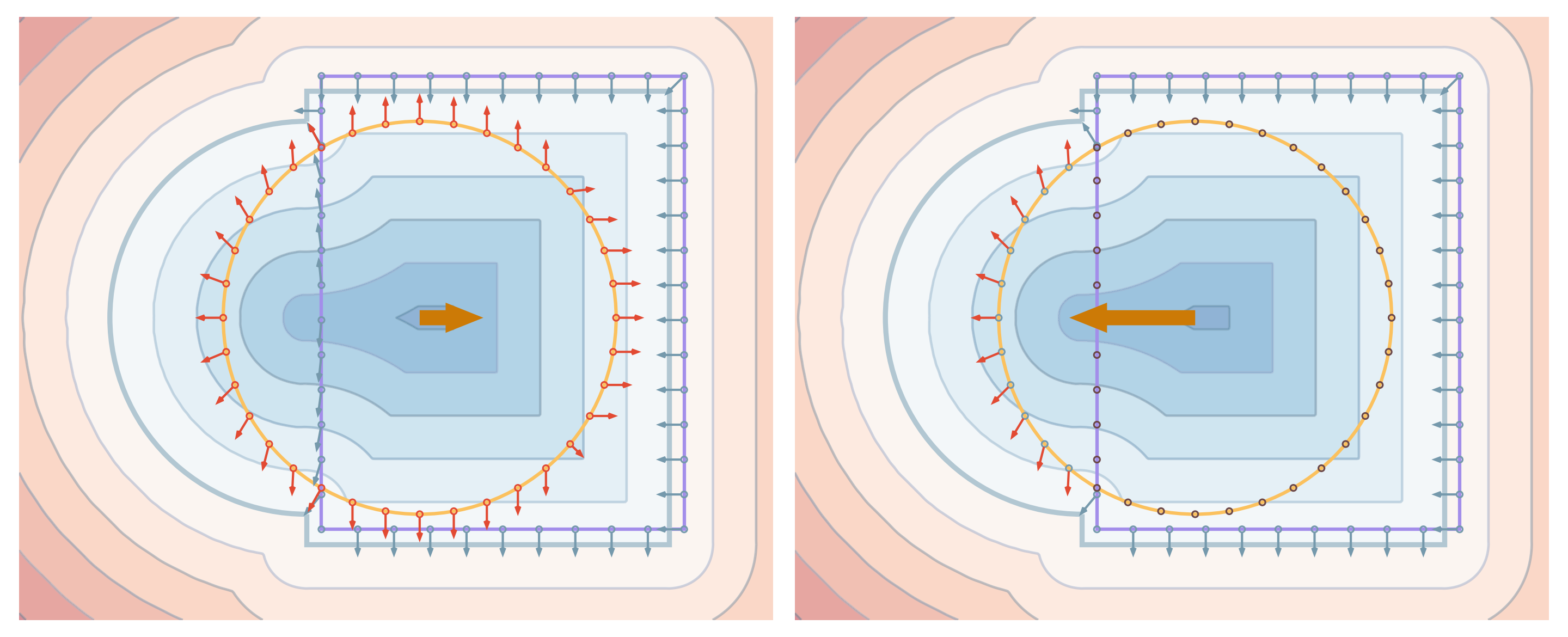}

    \caption{
    Visualization of the gradient in overlapping regions for a 2D toy case. By masking out the gradients from sampling points in the overlapping regions (indicated by the small red arrows within the rectangle), the yellow circle follows the correct optimization direction (orange arrows), aligning with the target geometry.
    }
    \label{fig:experiment_gradient}
\end{figure}

\begin{table}
    \centering
    \caption{
An ablation study examining the impact of temporal attention and the stopping of overlapped point gradients. The evaluation is based on metrics for part trajectory tracking, which assess both the temporal consistency of the generated exploded dynamics and the accuracy of part trajectory tracking. Enabling temporal attention improves the temporal consistency of the generated dynamics, while stopping gradients for overlapping points enhances the accuracy of part trajectory tracking.
}
    \begin{tabular}{ccc}
    \hline\hline
    \textbf{Variants}& \textbf{Weighted IoU} $\uparrow$ & \textbf{SDF Objective} $\downarrow$  \\\hline\hline
    w/o temporal attention   & $0.6874$& $0.0124$ \\
    w/o stopping gradients  & $0.7665$& $0.0092$ \\
    ours full  & $\bm{0.8163}$ & $\bm{0.0085}$  \\
    \hline
    \end{tabular}
    \label{tab:ablation}
\end{table}

\paragraph{Evaluation of Temporal Attention}
We conduct an ablation study to evaluate the effectiveness of the temporal attention mechanism in improving the quality of generated sequences. As shown in Table~\ref{tab:ablation}, incorporating temporal attention leads to a significant improvement in both metrics: a $18.8\%$ increase in weighted IoU and a $31.5\%$ reduction in the SDF objective. This demonstrates that temporal attention enhances temporal consistency and explosive linearity by enabling tokens to share information across different frames, ensuring smoother and more accurate part movements.

\paragraph{Evaluation of Stopping Overlapped Point Gradients}
Next, we investigate the impact of our method for stopping overlapped point gradients. In Fig.~\ref{fig:experiment_gradient}, we visualize a 2D example where overlapping parts can lead to incorrect gradient directions during optimization. In this example, the cyan contours represent the target geometry boundaries, with the SDF values shown in the background. The optimization of the yellow circle’s translation is considered, where the gradients contributing to the translation are depicted by the small red arrows.
When using uniform surface point sampling across the entire object, gradients from overlapping regions contribute equally, resulting in incorrect optimization directions (orange arrows) for the yellow circle. Our method resolves this issue by masking out the gradients from sampling points within the overlapped regions, leading to correct optimization directions (orange arrows) that align with the target geometry. This adjustment ensures that the optimization process is not adversely affected by overlapping regions, which is crucial in real-world 3D modeling where parts often overlap.
As shown in Table~\ref{tab:ablation}, incorporating this technique significantly improves the fitting metric, demonstrating that our method addresses the challenges posed by overlapping components and enhances the overall accuracy of the part fitting process.

\begin{figure}
    \centering
    \includegraphics[width=1\linewidth]{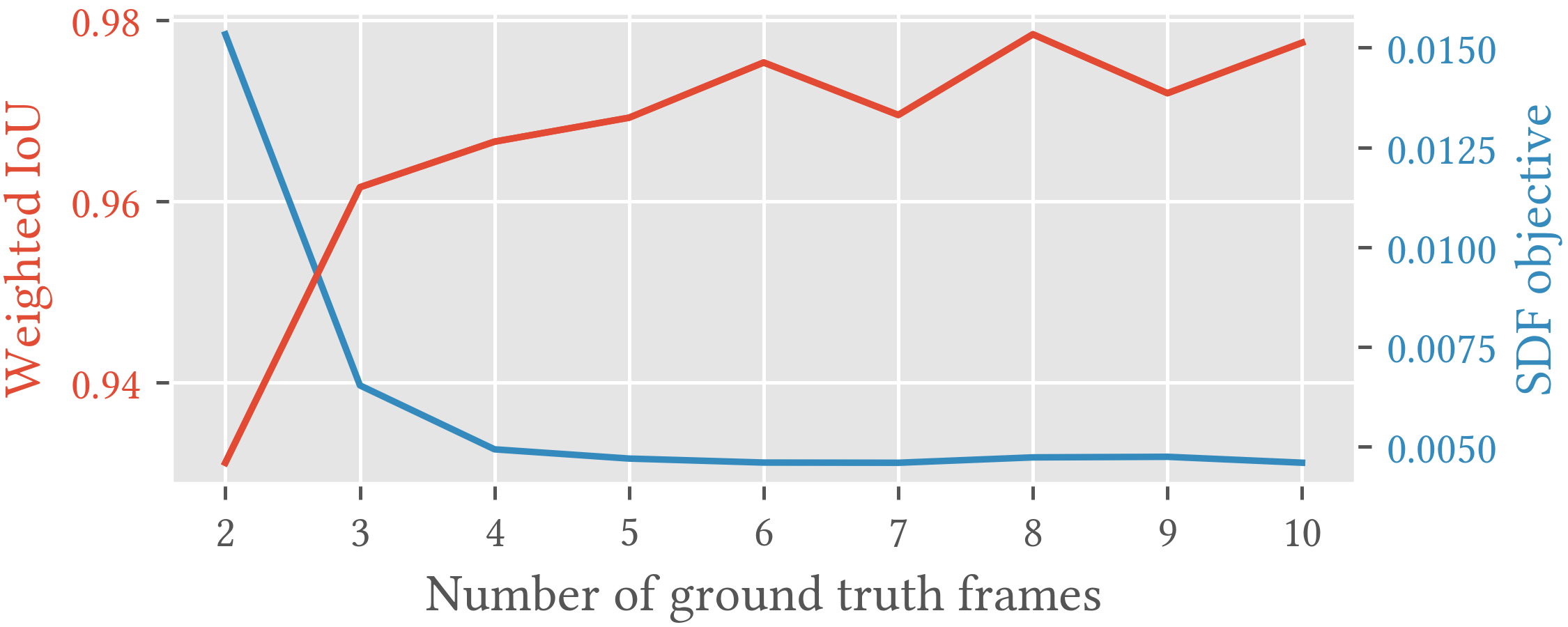}\\
    \vspace{0.25cm}
    \includegraphics[width=1\linewidth]{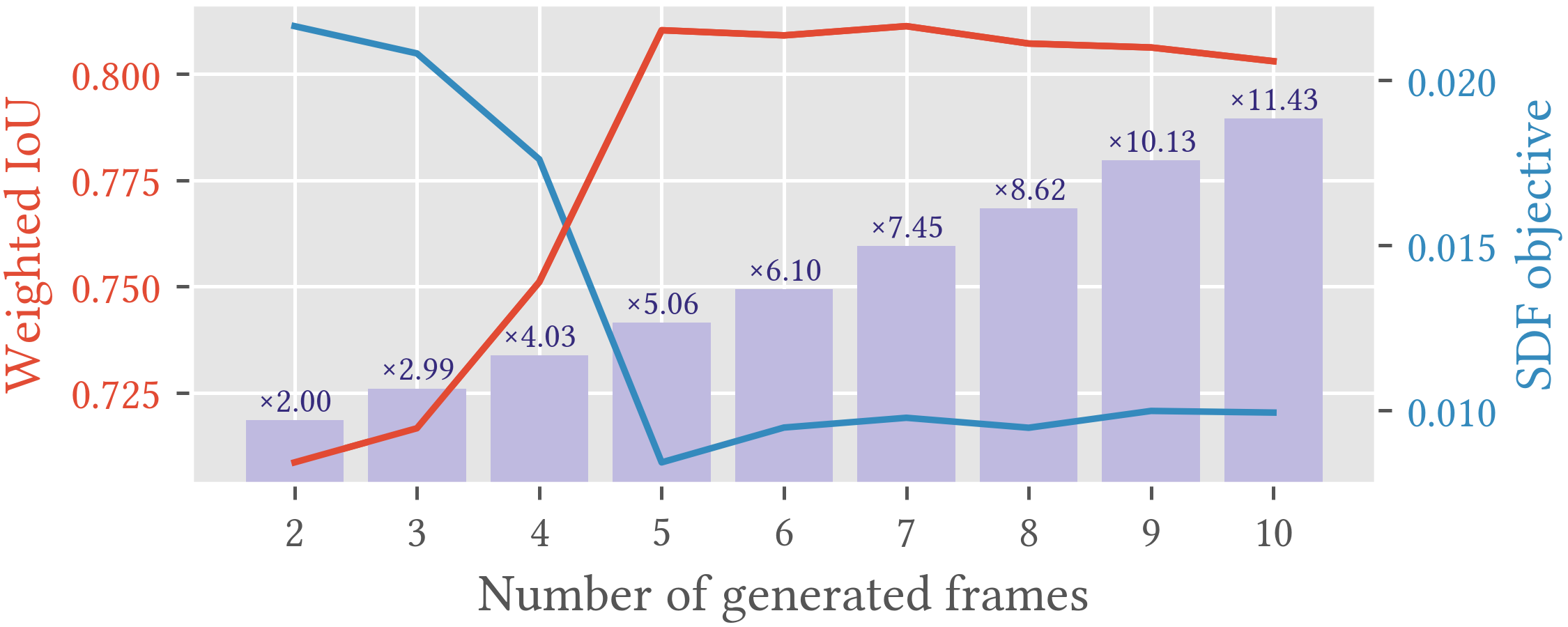}\\

    \caption{
Quantitative analysis of the impact of frame number on part trajectory tracking.
Top: For ground truth sequences, both weighted IoU (red) and SDF objective (blue) stabilize after 3 frames, indicating that more frames improve tracking accuracy.
Bottom: For generated exploded dynamics, performance metrics continue to improve up to 5 frames, while computational time cost (purple bars) increases with more frames. Although our model was trained with a maximum of 5 frames due to GPU memory limitations, this result suggests that 5 frames offer a reasonable trade-off between tracking accuracy and computational efficiency, balancing performance with processing time.
    }
    \label{fig:experiment_frames}
\end{figure}

\paragraph{Evaluation of Number of Frames Generated}
To evaluate the impact of sequence length on the quality of generated exploded dynamics, we analyze part trajectory tracking across varying numbers of input frames, using both ground truth (synthetic) and generated sequences. As shown in Fig.~\ref{fig:experiment_frames}, for ground truth sequences, both quality metrics—weighted IoU and SDF objective—show rapid convergence with just 3 frames, indicating that the explosion dynamics can be effectively captured with minimal temporal sampling. This curve suggests that increasing the number of frames improves the tracking accuracy.
For generated sequences, the metrics continue to improve until 5 frames. During training, our network was only trained with up to 5 frames given the constraints of training and GPU memory limitations, so performance naturally starts to drop after that point. However, the results still show some generalizability beyond the 5-frame training limit, demonstrating that more frames improve the accuracy of part trajectory tracking. 
As illustrated in the figure, the computational cost increases at a rate slightly faster than linear as the number of frames increases, making it impractical to use excessively long sequences due to the high inference time and memory requirements. Therefore, while more frames theoretically improve accuracy, a balance between training, inference time, and quality led us to select 5 frames as the optimal configuration.

\begin{figure}
    \centering
    \includegraphics[width=1\linewidth]{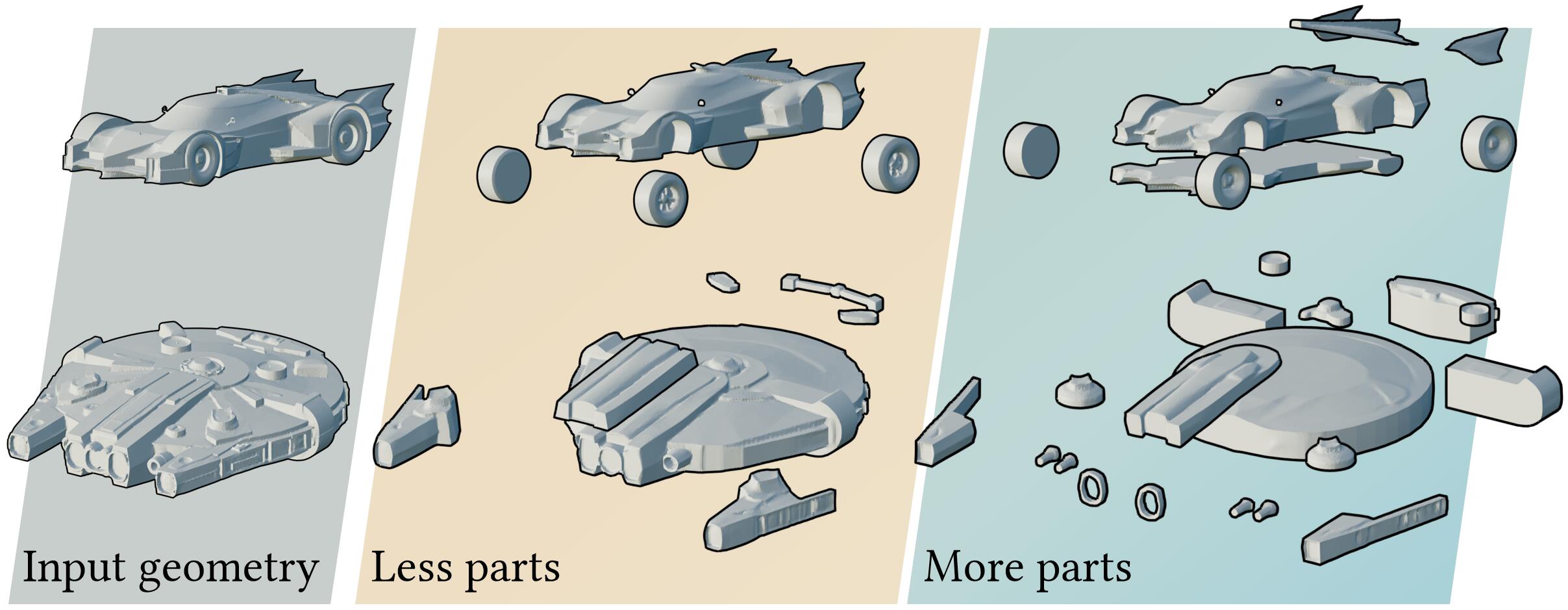}
    \caption{
Evaluation of the effect of part number embedding. The figure demonstrates the ability to control the number of parts through part number embedding. While achieving precise control can be challenging, the approach enables coarse control, where increasing the number of exploded parts in the embedding leads to the generation of more parts.
    }
    \label{fig:experiment_count}
\end{figure}

\paragraph{Evaluation of Part Number Control}

Our model allows control over the number of generated parts by adjusting the parts count during the generation process, as described in Sec.~\ref{sec:exploded_view_adapter}. While achieving precise control over the exact number of parts can be challenging—particularly for a diffusion model due to the continuous nature of the process—the model demonstrates the ability to adjust the number of exploded parts at a coarse level.
Fig.~\ref{fig:experiment_count} shows the results of controlling the number of exploded parts for the same input geometry. The model effectively adjusts the segmentation granularity, generating fewer parts when specified and more parts when a higher count is requested. This control is achieved without compromising the semantic consistency of the object. For example, the model merges functionally related components when fewer parts are specified, while a more detailed structural decomposition is produced when more parts are generated. These results confirm that our method strikes a balance between controlling segmentation granularity and maintaining semantic coherence.

\section{Discussions and Conclusions}

In this work, we introduce BANG, a generative framework that dynamically decomposes complex 3D assets into interpretable part-level structures via a smooth and consistent exploded view process. Built on a large-scale 3D generative model, BANG integrates two core components: the Exploded View Adapter, which conditions the model on input geometry and timestamps, and the Temporal Attention Module, which ensures smooth transitions across the exploded process. This framework captures sophisticated structural insights, enabling high-quality 3D decomposition, generation, and enhancement. BANG seamlessly integrates part-level multimodal analysis into creative workflows, making it a versatile tool for enhancing digital creation, especially where intuitive, component-based design is crucial. By mimicking the natural process of deconstruction and reassembly, BANG not only advances current 3D technologies but also aligns with human cognitive processes of understanding and creativity. Future work focused on improving physical realism, incorporating material properties, and expanding applicability could significantly enhance its potential, empowering creators across industries to bring complex designs to life.

\begin{figure}
    \centering
    
    \includegraphics[width=1\linewidth]{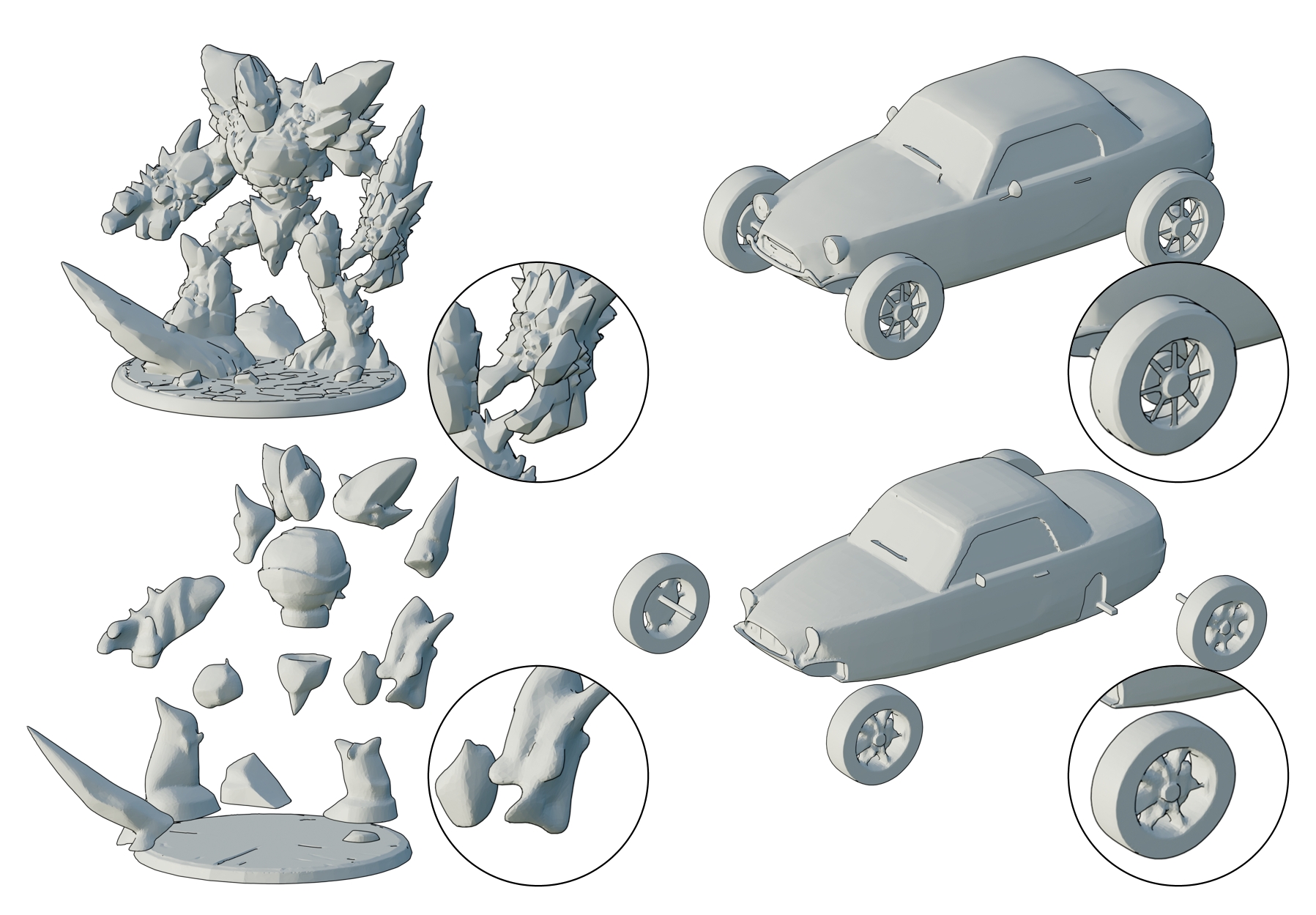}

    \caption{
Failure cases on highly detail geometry. Top: input meshes with complex structures. Bottom: generated exploded views. While BANG captures the overall structure, local detail is lost, and the exploded geometry drifts from the original. This is due to the lack of per-part supervision and limited token length in the current latent representation.
}
    \label{fig:failure}
\end{figure}
\paragraph{Limitations and Future Work.} 
Despite its strengths, BANG faces several limitations. While trained on 20k exploded dynamic data, it struggles with highly complex objects, particularly those with poorly defined structural components. Expanding the dataset to include a wider range of intricate structures, particularly real-world mechanisms, is essential for improving BANG's ability to handle more diverse 3D assets.
Another challenge is the preservation of precise geometric details during the exploded dynamics generation process. Although BANG isolates and regenerates parts at a high level of detail, subtle discrepancies between the exploded views and the original geometry persist, and some local details are lost in the process.
As illustrated in Fig.~\ref{fig:failure}, the generated exploded views exhibit noticeable deviation from the original geometry, particularly in highly detailed regions. This is due to the lack of explicit per-part geometric supervision during training, and the limited latent token length, which constrains the model’s ability to represent detailed geometry at part level.
Future research could incorporate advanced geometric constraints and scale up the model training to minimize these discrepancies, ensuring that exploded geometry aligns more closely with the original geometry while benefiting from part-level regeneration.
Currently, BANG follows an artistic pipeline tailored for visual representation, which may not fully meet the needs of applications that require realistic mechanical assembly or physical constraints, such as in manufacturing or robotics. While effective for digital design and visualization, bridging the gap between artistic modeling and engineering realism is necessary for industrial applications. Future versions could incorporate physical simulation techniques to account for material properties, structural interactions, and real-world assembly processes.
Finally, BANG currently focuses exclusively on geometry, neglecting material properties (e.g., flexibility, weight distribution, or compatibility) as well as appearance attributes (e.g., color or texture). Material and appearance considerations both play crucial roles in real-world assembly and disassembly tasks, affecting not only how parts physically interact and fit together but also how they are visually perceived. Integrating material properties alongside appearance attributes into BANG could improve its ability to handle realistic disassembly tasks, particularly in fields like product teardown, repair, manufacturing, and design, where these factors strongly influence the process.

\begin{acks}

This work was supported by National Key R\&D Program of China (2022YFF0902301), NSFC programs (61976138, 61977047), STCSM (2015F0203-000-06), and SHMEC (2019-01-07-00-01-E00003). We also acknowledge support from Shanghai Frontiers Science Center of Human-centered Artificial Intelligence (ShangHAI), MoE Key Lab of Intelligent Perception and Human-Machine Collaboration (ShanghaiTech University), Core Facility Platform of Computer Science and Communication of ShanghaiTech University, and HPC Platform of ShanghaiTech University.

\end{acks}

\bibliographystyle{ACM-Reference-Format}
\bibliography{sample-bibliography}

\end{document}